\documentclass[11pt,a4paper,onecolumn]{article}

\usepackage[T1]{fontenc}
\usepackage[utf8]{inputenc}
\usepackage{lmodern}
\usepackage[margin=1in]{geometry}
\usepackage{amsmath,amssymb,amsthm,mathtools,bm,amsfonts}
\usepackage{graphicx}
\usepackage{xcolor}
\usepackage{booktabs}
\usepackage{algorithm}
\usepackage{algpseudocode}
\usepackage{enumitem}
\usepackage{subcaption}
\usepackage{multirow}
\usepackage{tabularx}
\usepackage{array}
\usepackage{tikz-cd}
\usepackage{empheq}
\usepackage[numbers,square,sort&compress]{natbib}
\usepackage[hidelinks]{hyperref}
\graphicspath{{figures/}}

\providecommand{\Description}[1]{}

\newcommand{\U}{\mathcal{U}}
\newcommand{\E}{\mathbb{E}}
\newcommand{\R}{\mathbb{R}}

\newcommand{\cP}{\mathcal{P}}

\newcommand{\cD}{\mathcal{D}}

\newcommand{\rr}{\bm{r}}
\newcommand{\pp}{\bm{p}}
\newcommand{\C}{\mathcal{C}}
\newcommand{\Sl}{\mathcal{S}}
\newcommand{\Q}{\mathcal{Q}}
\newcommand{\setd}{{ d \kern -.15em l}}
\newcommand{\hatsetd}{ d \hat{\kern -.15em l }}
\newcommand{\dd}{\mathsf{d\kern -0.07em l}}
\newcommand{\X}{\mathcal{X}}

\newtheorem{definition}{Definition}

\newtheorem{proposition}{Proposition}

\newcommand{\tightparagraph}[1]{%
  \par\addvspace{0.25\baselineskip}%
  \textbf{\textit{#1.}}\enspace\ignorespaces
}

\allowdisplaybreaks[4]

\begin{document}

\title{
IncVNM: An Interactive Recommender System with Incremental 
von Neumann–Morgenstern
Utility Learning 
}

\author{Jiaxin Wei\thanks{School of Mathematics and Statistics, Xi’an Jiaotong University, Xi’an, Shaanxi, P. R. China, \texttt{weijx0609@stu.xjtu.edu.cn}}, 
Jia Liu\thanks{School of Mathematics and Statistics, Xi’an Jiaotong University, Xi’an, Shaanxi, P. R. China, \texttt{jialiu@xjtu.edu.cn}}}
\date{}

\maketitle

\begin{abstract}
In modern recommender systems, users are often presented with products
whose outcomes are random. The values of such products depend on their entire outcome distributions and on users' attitudes toward risk, making them difficult to evaluate using traditional product-level recommendation methods. We propose \textsc{IncVNM}, an interactive recommender system that incrementally learns a user-specific von Neumann--Morgenstern (VNM) utility function from ordinal click feedback generated during ordinary webpage browsing. At each interaction, \textsc{IncVNM} converts the click order within a displayed product list into pairwise comparisons between outcome distributions and updates the nominal utility by minimizing its deviation from the previous estimate while incorporating the newly observed pairwise comparisons. To close the interaction loop, an entropy-regularized sampling policy balances exploitation of the learned utility against exploration for additional preference information. We evaluate \textsc{IncVNM} in a rolling-window experiment with different response-error probabilities and show that it substantially outperforms the expected-return, mean--variance, discounted-UCB, and distribution-aware BPR baselines.

\end{abstract}



\section{Introduction}
Recommender systems have become an essential component of online
platforms by learning user preferences from clicks, purchases, and
browsing behaviors. Existing approaches, including collaborative
filtering \citep{resnick1994grouplens}, sequential recommendation models, and deep representation
learning \citep{hidasi2016parallel}, mainly capture statistical associations between users and
items. However, many products have inherently random outcomes, such
as financial assets with volatile returns, insurance contracts with
contingent payoffs, and services with uncertain costs. Such products are
characterized by outcome distributions rather than deterministic features. Moreover, the outcome distributions may evolve over time, while users evaluate these products dynamically according to their individual risk
attitudes. Consequently, product-level preference scores learned from
historical interactions may perform poorly in dynamic or transfer settings
as the outcome distributions
change.

The von Neumann--Morgenstern (VNM) utility function naturally provides
an interpretable and transferable representation of user preferences
over random outcomes. By learning this function from user feedback,
the recommender system captures preferences over outcomes rather than
specific product features. The learned utility can therefore be
applied to new products and to existing products with time-varying outcome
distributions. This enables preference evaluation based on current
distributions and mitigates temporal transferability problems without
requiring new product-specific interaction histories. However,
re-learning the utility function from the entire interaction history
would impose a substantial computational burden in an online environment where feedback arrives continuously.

To address this issue, we propose {IncVNM}, an interactive
recommender system that \emph{incrementally learns} user-specific VNM utility
functions from streaming click sequences. Rather than relearning the
utility function from scratch, IncVNM updates the current estimate after
each interaction using newly extracted preference information and the
previous estimate.
The dimension of each per-user update
is independent of the product-catalog size and remains uniformly bounded
over the interaction history. This enables lightweight streaming updates
and immediate evaluation of both new products and products with updated
outcome distributions under the finite-precision
representation.
To complete the interactive recommendation loop, IncVNM adopts an
entropy-regularized sampling policy that balances exploration and
exploitation. The interactive framework is summarized as follows:
\[
\begin{tikzcd}[column sep=2em, row sep=2em]
\text{product list generation}
  \arrow[r]
&
\text{click sequence}
  \arrow[d]
\\
\text{utility update}
  \arrow[u]
&
\text{pairwise comparisons}
  \arrow[l]
\end{tikzcd}
\]


Our main contributions are summarized as follows:
\begin{itemize}[leftmargin=*]
    \item \textbf{Interactive VNM recommendation:} We formulate an
interactive recommender system for products with random outcomes from click sequences by representing user preferences through a VNM utility function. 

    \item \textbf{Incremental learning framework:} We propose IncVNM, which incrementally updates a utility function through a tractable optimization model. Each update is independent of the catalog size and remains uniformly bounded over time.

    \item \textbf{Unified learning--recommendation framework:}
We integrate explicit preference learning and product recommendation into
a unified interactive framework, in which the learned preferences are
immediately used to generate subsequent recommendations. Numerical
experiments demonstrate superior utility-learning and recommendation
performance over four baselines. 
\end{itemize}

 The remainder of the paper is organized as follows. 
 Section~\ref{sec:pre-clicit} introduces the recommender system
 and the incremental VNM utility-updating method. Section~\ref{sec:rec-alg} presents the entropy-regularized product-list generation policy and the 
 IncVNM algorithm. Section~\ref{sec:numerical} reports the numerical 
 experiments.
 Section~\ref{sec:conclusion} concludes.

\section{Related work}\label{sec:related}
\tightparagraph{\textbf{Recommender systems}}
Recommender systems have evolved from classical content-based and collaborative filtering \citep{adomavicius2005toward,pazzani2007content,resnick1994grouplens,linden2003amazon} to matrix factorization and low-dimensional latent space representations \citep{koren2009matrix}. The field subsequently shifted toward implicit behavioral feedback and ranking-oriented learning, incorporating pairwise ranking and contextual bandits \citep{rendle2009bpr,li2010contextual}. Modern architectures further advance personalization through deep sequential models (e.g., GRU4Rec, SASRec, BERT4Rec) \citep{hidasi2016parallel,kang2018self,sun2019bert4rec}, graph neural networks \citep{he2020lightgcn}, and generative retrieval \citep{rajput2023recommender}.

Despite these advances, mainstream recommender systems primarily learn statistical user--item associations rather than interpretable preference representations. They generally do not represent product uncertainty at the level of outcome distributions or account for individual risk attitudes, thereby adapting poorly to time-varying random outcomes.
These limitations motivate learning a transferable, utility-function-level
risk profile from sequential behavioral feedback.

\tightparagraph{Preference modeling and elicitation for products with random outcomes}
\label{subsec:related_utility_elicitation}
Products with random outcomes cannot be fully evaluated through fixed features, as their value depends on both outcome distributions and individual risk attitudes. 
The random outcome of products can be evaluated via VNM expected utility \cite{von1944theory}. Classical studies elicited VNM utility statically through pre-designed lottery questionnaires and gamble tradeoffs \citep{farquhar1984state,wakker1996eliciting}. Subsequent work infers VNM utility from discrete choice data using parametric or non-parametric statistical choice models \citep{eliashberg1985measurement,hey1994investigating,chen2026eliciting}. Complementary optimization-based approaches maintain an ambiguity set of admissible utility functions or use polyhedral adaptive queries to reduce preference ambiguity \citep{armbruster2015decision,hu2015robust,guo2024utility,zhang2025modified}.

Despite their theoretical rigor, existing VNM elicitation methods are typically offline and query-based, requiring explicit user input beyond ordinary platform interactions. Their batch optimization formulations are also not perfectly 
suitable for real-time streaming updates. These limitations motivate us to integrate elicitation and recommendation into a single interactive loop, incrementally learning VNM utility from implicit click sequences through a lightweight optimization problem.

\section{Preference elicitation from click sequences}\label{sec:pre-clicit}
\subsection{Pairwise comparisons from click sequences}
\begin{figure}[htbp]
    \centering
    \includegraphics[width=0.8\linewidth]{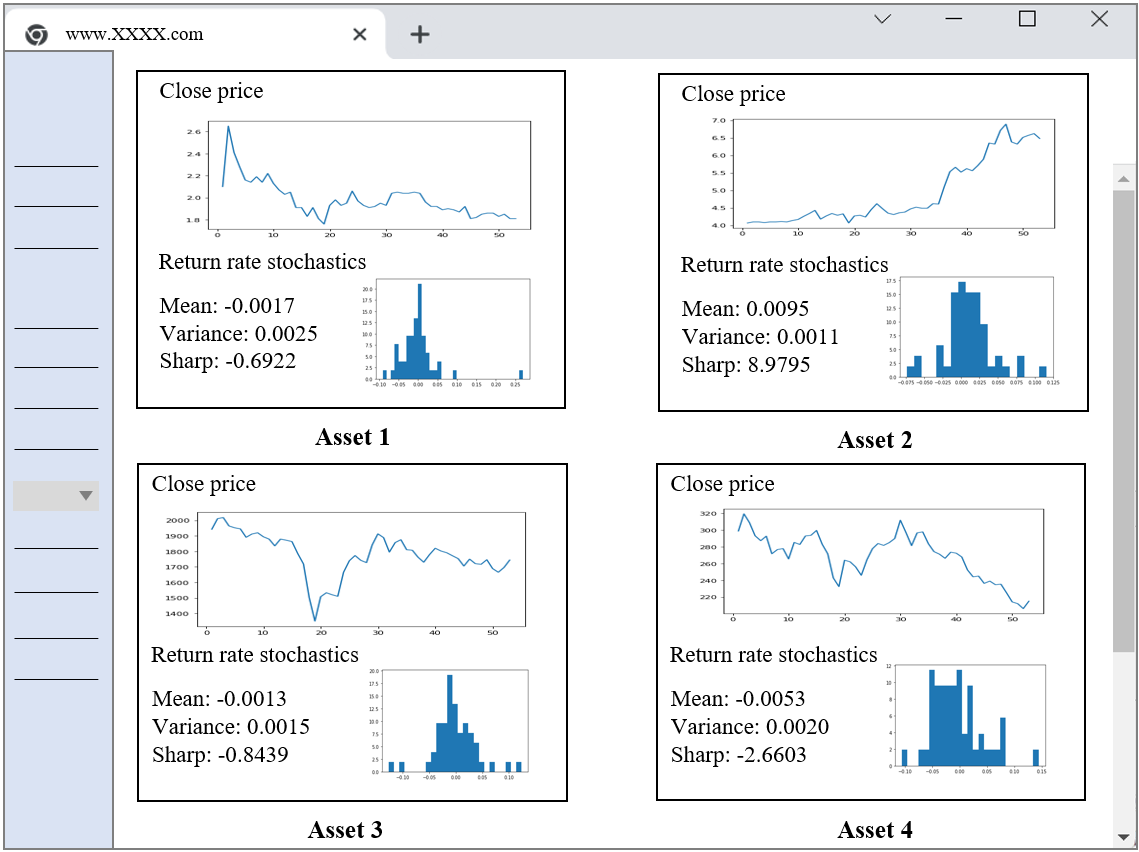}
    \caption{Illustrative interface for displaying random-outcome products.}
    \Description{A browser-style interface displaying four product cards, each containing historical outcome information and summary statistics.}
    \label{fig:interface}
\end{figure}

Nowadays, users commonly browse online platforms to search for product information before making a purchase decision. After a search, the platform displays $n$ products on the screen, referred to as a \emph{product list} and denoted by
$
\Sl=(S_1,\ldots,S_n);
$
see Figure~\ref{fig:interface} for an illustrative example.

During ordinary webpage browsing, a user may first click the product that attracts the most interest, return to the product list, and then click the next most preferred product among the remaining alternatives. This process continues until none of the remaining products attracts sufficient interest, after which the user may scroll down to view a new product list. The resulting ordinal feedback is referred to as a \emph{click sequence}.

For a displayed product list $\Sl$, let
$
\C_{\Sl}=(C_1,\ldots,C_T),
\; 1\leq T\leq n,
$
denote the corresponding click sequence, where $C_t\in\Sl$ is the product selected at the $t$-th click and
$
\{C_1,\ldots,C_T\}\subseteq\Sl.
$
We assume that each product is clicked at most once.

\begin{figure}[t]
    \centering
    \includegraphics[width=0.5\linewidth]{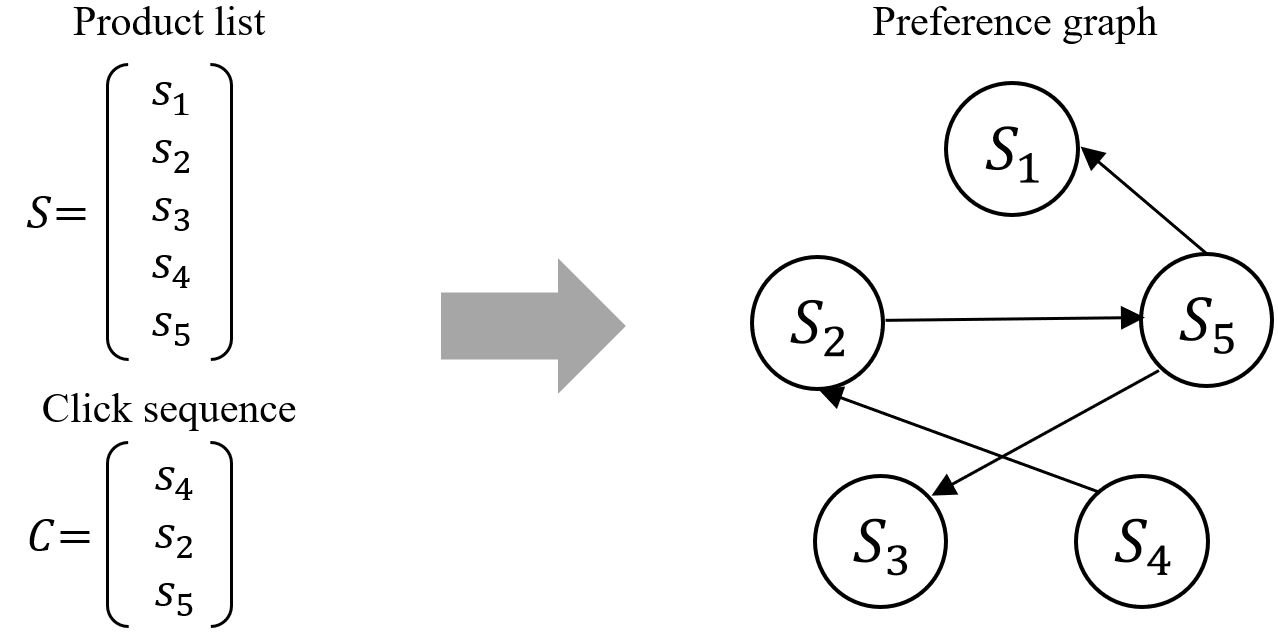}
    \caption{Pairwise comparisons from a click sequence.}
    \Description{A click sequence over five products is converted into a directed preference graph representing the induced pairwise comparisons.}
    \label{fig:click-example}
\end{figure}

We interpret click sequences as ordinal preference signals, where earlier clicks provide stronger evidence of preference, then a product list and its corresponding click sequence induce the following preference relations:
\begin{subequations}\label{eq:preference-order}
\begin{align}
    &C_i \succeq C_{i+1},
     \quad\text{for}\; i=1,\ldots,T-1,     \label{eq:click-order}\\
    &C_T \succeq S_j,
    \quad\;\;\text{for}\; S_j\in\Sl\setminus\{C_1,\ldots,C_T\}.
\label{eq:pre-order-rest}
\end{align}
\end{subequations}
Relation~\eqref{eq:click-order} states that a product clicked earlier is  preferred to one clicked later. Relation~\eqref{eq:pre-order-rest} states that the last clicked product is preferred to every unclicked product in the product list. 
Figure~\ref{fig:click-example} provides a simple illustration.
For notational convenience, we collect all pairwise comparisons induced by $(\Sl,\C_{\Sl})$ into
\begin{equation}\label{eq:def-paircom}
\begin{aligned}
\Q(\Sl,\C_{\Sl})
:={}&
\{(C_i,C_{i+1})\mid i=1,\ldots,T-1\}
\\
&\cup
\{(C_T,S_j)\mid
S_j\in\Sl\setminus\{C_1,\ldots,C_T\}\}
\\
=:{}&
\{(A_\ell,B_\ell)\}_{\ell=1}^{n-1}.
\end{aligned}
\end{equation}
The click sequence generates $T-1$ comparisons between consecutively clicked products and $n-T$ comparisons between the last clicked product and the unclicked products, hence ${n-1}$ pairs in total. For each $(A_\ell,B_\ell)\in\Q(\Sl,\C_{\Sl})$, 
$A_\ell$ is observed to be preferred to $B_\ell$, namely, 
$A_\ell\succeq B_\ell$. If no product is clicked, set $\Q(\Sl,\C_{\Sl})=\emptyset$.

The system can then use the collected pairwise preference information to learn a user-specific utility function.

\subsection{Preference elicitation model}
\label{sec:pre-eli-mod}

After collecting a click sequence, the system uses the induced pairwise comparisons to elicit the user's preference. In this paper, we focus on user preferences over products with random outcomes.

\tightparagraph{Random-outcome products}
A random-outcome product is a product whose realized outcome is not known at the time of recommendation. Unlike deterministic products, \emph{a random-outcome product can not be evaluated through a fixed vector of features}, since it may generate different outcomes with different probabilities. Examples include the future return of a financial asset, the realized payoff of an insurance contract, and the eventual transaction cost of an airline ticket. The value of such a product therefore depends on the distribution of its possible outcomes, which may be \emph{time-varying}.

Let
$
\cD=\{A_1,\ldots,A_{|\cD|}\}
$
be the catalog of all products available for recommendation. Formally, each random-outcome product $A\in\cD$ is represented by a finite
outcome vector
$
\rr^A=(r_1^A,\ldots,r_{n_A}^A)
$
and the associated probability vector
$
\pp^A=(p_1^A,\ldots,p_{n_A}^A),
$
where
$
p_i^A\geq 0,
\;\text{and}\;
\sum_{i=1}^{n_A}p_i^A=1.
$
That is,
\begin{equation}\label{def:product}
    \mathbb{P}(A=r_i^A)=p_i^A,
    \quad\text{for}\; i=1,\ldots,n_A.
\end{equation}
The user evaluates product $A$ through its outcome distribution.

\tightparagraph{VNM utility function}
For random-outcome products, von Neumann--Morgenstern (VNM) expected utility theory provides a natural framework for representing user preferences. We therefore model the user's preferences over random-outcome products through VNM expected utility.

Let $[\underline{x},\overline{x}]$ denote the interval that contains possible outcomes of all products in $\cD$. We define $\mathcal{U}_L$ as the set of endpoint-normalized, non-decreasing, and $L$-Lipschitz continuous utility functions:
$$
\begin{aligned}
\mathcal{U}_L
:=
\Big\{
u \ \Big| \ &
u(\underline{x})=a,\;
u(\overline{x})=b,\;
u \text{ is non-decreasing},\\
&
|u(x)-u(y)|
\leq
L|x-y|,
\quad
\forall x,y\in[\underline{x},\overline{x}]
\Big\}.
\end{aligned}
$$
Here, $a<b$ are fixed normalization constants that remove the affine indeterminacy of the utility function without changing the induced preference ranking. For any utility function $u\in\mathcal{U}_L$, define the VNM expected-utility functional $V_u:\cD\to\mathbb{R}$ by
$$
V_u(A)
:=
\mathbb{E}[u(A)]
=
\sum_{i=1}^{n_A}
p_i^A u(r_i^A),
\qquad
\forall A\in\cD.
$$
This functional provides a distribution-level representation of product preferences based on their outcome distributions.

Given the observed pairwise comparisons, the system seeks a utility function $u\in\U_L$ whose induced expected-utility functional is consistent with the user's responses. Specifically, for each observed comparison $A\succeq B$, $u$ should satisfy
\begin{equation}
\label{eq:preference_compare}
V_u(A)
=
\sum_{i=1}^{n_A}
p_i^A u(r_i^A)
\geq
\sum_{i=1}^{n_B}
p_i^B u(r_i^B)
=
V_u(B).
\end{equation}
By introducing a VNM utility function, the observed preferences are represented in a common outcome space. The learned utility can therefore be used to evaluate newly introduced products 
or products with time-varying random outcomes. This utility-based preference representation is highly transferable, whereas product-specific scoring methods may fail when product outcome distributions change.


\tightparagraph{Utility updating}
Upon observing a nonempty click sequence, the system updates its estimate of the user's utility function through an \emph{incremental learning} approach. The updated utility should deviate as little as possible from the current estimate, thereby retaining previously learned information while incorporating the newly collected preference information.

At interaction round $k$, the system holds a nominal utility function $\tilde{u}^{k-1}$ learned from the previous $k-1$ rounds. It displays a product list
$
\Sl^k=(S_1^k,\ldots,S_n^k)\subseteq\cD
$
and observes the corresponding click sequence
$
\C^k=(C_1^k,\ldots,C_{T^k}^k),
$
where $C_t^k\in\Sl^k$ for $t=1,\ldots,T^k$. The pair $(\Sl^k,\C^k)$ induces the set of pairwise comparisons $\Q^k:=\Q(\Sl^k,\C^k)$
as defined in \eqref{eq:def-paircom}.

The system then updates the nominal utility by solving
\begin{subequations}\label{eq:updating}
\begin{eqnarray}
\tilde{u}^k\in & \mathop{\arg\min}\limits_{u\in\U_L}
&\dd_K(u,\tilde{u}^{k-1})
\label{eq:updating-obj}\\
& \text{s.t.}
&\E[u(A)]\geq\E[u(B)],
\ \ 
\forall (A,B)\in\Q^k.
\label{eq:hard_com}
\end{eqnarray}
\end{subequations}
Here, $\dd_K:\U_L\times\U_L\to\R_+$ denotes the Kantorovich distance between two utility functions, whose definition is in Appendix~\ref{apx:kanto}.

\tightparagraph{Piecewise-linear approximation (PLA)}
Since the utility updating problem \eqref{eq:updating} is a non-parametric optimization problem over functional space, it is difficult to solve directly. We therefore approximate $u$ by a piecewise-linear utility function to obtain a tractable finite-dimensional approximation.

Let
$
\X:=\{x_1,\ldots,x_{N+1}\},
\;
\underline{x}=x_1<\cdots<x_{N+1}=\overline{x},
$
be a prescribed set of breakpoints. Define
$$
\U_{\X}
:=
\left\{
u_{\X}\in\U_L
\ \middle|\
u_{\X}\text{ is affine on each }[x_i,x_{i+1}],
\ i=1,\ldots,N
\right\}\subseteq\U_L
$$
as the set of admissible piecewise-linear utility functions associated with the breakpoint set $\X$.
For any $u_{\X}\in\U_{\X}$, let
$$
\beta_i
:=
\frac{u_{\X}(x_{i+1})-u_{\X}(x_i)}
{x_{i+1}-x_i},
\quad
i=1,\ldots,N,
$$
denote its slopes on each interval. Then $u_{\X}$ can be represented as
\begin{equation}\label{eq:def_PLU}
u_{\X}(x)
=
u_{\X}(x_i)+\beta_i(x-x_i)
\;\;
\text{for}\;x\in[x_i,x_{i+1}],
\;
i=1,\ldots,N.
\end{equation}
The endpoint normalization requires
$
u_{\X}(x_1)=a\;\text{and}\;
u_{\X}(x_{N+1})=b.
$
Therefore, once the breakpoint set $\X$ is specified, a normalized piecewise-linear utility function is uniquely determined by its slope vector
$
\bm{\beta}=(\beta_1,\ldots,\beta_N).
$

For any utility function $u\in\U_L$, we define its \emph{piecewise-linear approximation} $u_{\X}\in\U_{\X}$ by linear interpolation:
\begin{align}\label{eq:def_PLA}
(\Pi_{\X}u)(x)
=
u(x_i)
&+
\frac{u(x_{i+1})-u(x_i)}
{x_{i+1}-x_i}
(x-x_i),\nonumber
\\
&\text{for }x\in[x_i,x_{i+1}],
\quad
i=1,\ldots,N.
\end{align}
In particular,
$
(\Pi_{\X}u)(x_i)=u(x_i),
\;
i=1,\ldots,N+1
$ and $
\Pi_{\X}u\in\U_{\X}$.

By approximating the utility function $u$ in problem \eqref{eq:updating} with a piecewise-linear utility function $u_{\X}$, we obtain the following approximate utility-updating problem:
\begin{subequations}\label{eq:updating_PLA}
\begin{eqnarray}
\tilde{u}_{\X}^{k}
\in
& \mathop{\arg\min}\limits_{u_{\X}
\in\U_{\X}}
&\dd_K
\left(
u_{\X},
\tilde{u}_{\X}^{k-1}
\right)
\label{eq:updating_PLA_obj}\\
&\text{s.t.}
&\E[u_{\X}(A)]\geq\E[u_{\X}(B)],
\quad
\forall (A,B)\in\Q^k.
\label{eq:updating_PLA_cons}
\end{eqnarray}
\end{subequations}
Here, $\tilde{u}_{\X}^{k-1}$ is the piecewise-linear nominal utility learned from the previous $k-1$ rounds of interaction. When problem \eqref{eq:updating_PLA} is represented in terms of the slope vector $\bm{\beta}$, it can be reformulated as a second-order cone program (SOCP) in \eqref{eq:final-reformulation}; see more details of the reformulation in Appendix~\ref{apx:reform_updating}.

\begin{figure}[!b]
    \centering
    \includegraphics[width=0.8\textwidth]{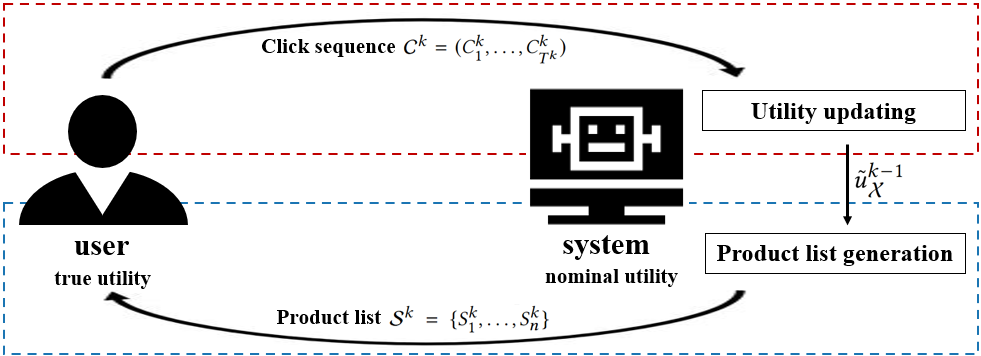}
    \caption{Overview of the IncVNM interaction loop. The utility-learning process highlighted in the red dashed box is described in Section~\ref{sec:pre-eli-mod}, while the product-list generation process highlighted in the blue dashed box is developed in Section~\ref{sec:product-ge}.}
    \Description{An overview of the IncVNM interaction loop, including product-list generation, user clicks, pairwise comparison extraction, and nominal utility updating.}
    \label{fig:system-framework}
\end{figure}

\tightparagraph{Relaxation for response errors}
In real-world interactions, click sequences may contain response errors caused by accidental clicks or difficulty in distinguishing similar products. Such errors may induce pairwise comparisons that
cannot be represented by any utility function in the admissible
utility class. 
Consequently, imposing all induced pairwise comparisons as hard constraints in \eqref{eq:updating_PLA_cons} may render the utility-updating problem infeasible. We address this issue by penalizing confidence-weighted violation.

First, to mitigate the loss of ordinal information when click sequences contain response errors, we extend the adjacent-click relations in \eqref{eq:click-order} to all pairs of clicked products:
\begin{equation}
\label{eq:click-order-relax}
C_i\succeq C_j,
\qquad
1\leq i<j\leq T.
\end{equation}
We denote the resulting extended set of pairwise comparisons induced by $(\Sl,\C_{\Sl})$ by
$
\tilde{\Q}(\Sl,\C_{\Sl}).
$
At interaction $k$, let
$
\tilde{\Q}^k
:=
\tilde{\Q}(\Sl^k,\C^k).
$

For each comparison
$
q=(A_q,B_q)\in\tilde{\Q}^k,
$
define its utility margin under $u\in\U_L$ as
$$
\Delta_u(q)
:=
V_u(A_q)-V_u(B_q)
=
\mathbb{E}[u(A_q)]
-
\mathbb{E}[u(B_q)].
$$
Then the corresponding 
preference constraint \eqref{eq:updating_PLA_cons} is
$
\Delta_{u_\X}(q)\geq 0.
$


Second, to accommodate potentially conflicting comparisons, we introduce a nonnegative violation variable $\xi_q$ for each
$
q\in\tilde{\Q}^k.
$
The relaxed utility-updating model is
\begin{subequations}
\label{eq:confidence_weighted_update}
\begin{eqnarray}
\tilde{u}_{\X}^{k}
\in
&
\mathop{\arg\min}\limits_{
{
u_{\X}\in\U_{\X}, 
\bm{\xi}\geq 0}}
&
\dd_K
\left(
u_{\X},
\tilde{u}_{\X}^{k-1}
\right)
+
\gamma
\sum_{q\in\tilde{\Q}^k}
\omega_q^k\xi_q
\label{eq:confidence_weighted_update_obj}
\\
&\text{s.t.}
&
\Delta_{u_{\X}}(q)+\xi_q\geq 0,
\quad
\forall q\in\tilde{\Q}^k.
\label{eq:confidence_weighted_preference}
\end{eqnarray}
\end{subequations}
For each
$
q\in\tilde{\Q}^k,
$
the confidence weight is defined as
\begin{equation}
\label{eq:confidence-weight}
\omega_q^k
:=
\frac{2}{
1+
\exp\left(
-m
\left|
\Delta_{\tilde{u}_{\X}^{k-1}}(q)
\right|
\right)
}
-1
\in[0,1),
\end{equation}
where $m>0$ is a scaling parameter. Motivated by sigmoid preference models for pairwise human feedback \citep{christiano2017deep}, the weight $\omega_q^k$ increases with the absolute nominal utility margin. Thus, comparisons between products that are more distinguishable under the current nominal utility receive larger violation penalties. The parameter $\gamma\geq 0$ controls the trade-off between preserving the previous nominal utility and fitting the newly observed comparisons. The violation variables guarantee feasibility even when the observed comparisons are mutually inconsistent.

\section{Product recommendation
and the IncVNM algorithm}
\label{sec:rec-alg}

Section \ref{sec:pre-clicit} describes how IncVNM converts the collected click sequence into pairwise comparisons and updates the nominal utility function, corresponding to the process highlighted in the red dashed box in Figure~\ref{fig:system-framework}. This section turns to the complementary step of recommending products (generating the product list) displayed in the next interaction based on the current nominal utility, corresponding to the process highlighted in the blue dashed box.

\subsection{Product-list generation with entropy regularization}\label{sec:product-ge}
Suppose that the system currently holds a nominal utility function $\tilde{u}$. When generating a product list (recommendation), \emph{exploitation} refers to recommending products with the highest nominal expected utilities, which may provide attractive short-term recommendations but limited information for further preference learning. By contrast, \emph{exploration} refers to 
exposing the user to more diverse products and collecting additional preference information. Exploration may therefore yield long-term benefits by improving preference learning, but it may reduce immediate recommendation quality. We therefore use entropy regularization to balance exploitation and exploration.

To generate a product list, the system samples $n$ products from $\cD$ at a time. Then, the key point is to determine the sampling probabilities. 
Let
$
\cP_\cD
:=
\left\{
\bm{\mu}\in\mathbb R_+^{|\cD|}
\ \middle|\
\sum_{i=1}^{|\cD|}\mu_i=1
\right\}
$
denote the probability simplex over $\cD$, where $\mu_i$ is the sampling probability assigned to product $A_i$. The sampling distribution is obtained by solving
\begin{equation}
\label{eq:entropy-policy}
\max_{\bm{\mu}\in\cP_\cD}
\quad
\sum_{i=1}^{|\cD|}
\mu_i V_{\tilde u}(A_i)
-
\lambda
\sum_{i=1}^{|\cD|}
\mu_i\log\mu_i.
\end{equation}
where $\lambda>0$ is the exploration rate controlling the strength of exploration. The first term favors products with high nominal expected utility, whereas the latter entropy term encourages a more dispersed sampling distribution.

The optimal solution to \eqref{eq:entropy-policy} has a softmax form
\begin{equation}
\label{eq:softmax}
\mu_i
=
\frac{
\exp\left(V_{\tilde u}(A_i)/\lambda\right)
}{
\sum_{j=1}^{|\cD|}
\exp\left(V_{\tilde u}(A_j)/\lambda\right)
},
\quad
i=1,\ldots,|\cD|.
\end{equation}
As $\lambda\to 0$, the distribution concentrates on products with the highest nominal expected utilities. As $\lambda\to\infty$, it approaches the uniform distribution over $\cD$. The complete product-list generation procedure is summarized in Algorithm~\ref{alg:Pro-list-gen}.

\begin{algorithm}
\caption{Entropy-regularized product-list generation}
\label{alg:Pro-list-gen}
\begin{algorithmic}[1]
    \State \textbf{Input:} Product-list size $n$, nominal utility $\tilde{u}$, product catalog $\cD$, and regularization parameter $\lambda$.
    \State Compute the sampling distribution $\bm{\mu}$ according to \eqref{eq:softmax}.
    \State Sample $n$ distinct products from $\cD$ without replacement according to $\bm{\mu}$. The sampling order determines the display order.
    \State \textbf{Output:} $\Sl=(S_1,\ldots,S_n)$.
\end{algorithmic}
\end{algorithm}


Now we are ready to 
give a complete interaction loop under a fixed breakpoint set, shown in Algorithm~\ref{alg:rec-alg-fixed}. Specifically, at each interaction, the system generates a product list, observes the click sequence, and then updates the nominal utility.

Note that Algorithm~\ref{alg:rec-alg-fixed} presents the interaction loop for the error-free setting. In the presence of response errors, it solves \eqref{eq:confidence_weighted_update} instead of \eqref{eq:updating_PLA} in line~7 of Algorithm~\ref{alg:rec-alg-fixed}, with $\tilde{\Q}^k$ defined in Section~\ref{sec:pre-eli-mod}.

\begin{algorithm}[htbp]
\caption{Fixed-breakpoint recommendation loop}
\label{alg:rec-alg-fixed}
\begin{algorithmic}[1]
    \State \textbf{Input:} Breakpoint set $\X$, initial nominal utility
    $\tilde{u}_{\X}^{0}\in\U_{\X}$, and total number of interaction
    rounds $K$.
    \For{$k=1,\ldots,K$}
        \State Generate $\Sl^k$ from
        $\tilde{u}_{\X}^{k-1}$ using
        Algorithm~\ref{alg:Pro-list-gen}. Observe $\C^k$.
        \If{$|\C^k|=0$}
            \State Set $\Q^k=\varnothing$ and
            $\tilde{u}_{\X}^{k}=\tilde{u}_{\X}^{k-1}$.
        \Else
            \State Construct $\Q^k=\Q(\Sl^k,\C^k)$. Update the nominal utility by solving
            \eqref{eq:updating_PLA} and denote the solution by
            $\tilde{u}_{\X}^{k}$.
        \EndIf
    \EndFor
    \State \textbf{Output:} Final nominal utility
    $\tilde{u}_{\X}^{K}$.
\end{algorithmic}
\end{algorithm}


\subsection{Dynamic refinement of the 
breakpoint set}
Keeping the breakpoint set $\X$ fixed leads to two limitations.
First, it imposes an irreducible approximation error. For any
$u\in\U_L$, its PLA satisfies
\citep{wei2026coordinate,guo2024utility}
$$
\|u-\Pi_{\X}u\|_\infty
\leq
\frac{L}{4}h_{\X},
\qquad
h_{\X}
:=
\max_{i=1,\ldots,N}(x_{i+1}-x_i).
$$
Since this bound is determined by the mesh size $h_{\X}$, collecting
additional preference information cannot improve the approximation
unless the breakpoint set is refined. Hence, if the underlying utility
is not piecewise linear over $\X$, the approximation error cannot
vanish.
Second, a fixed breakpoint set may fail to preserve preference
directions. Let
$
u_{\X}^*:=\Pi_{\X}u^*
$
be the PLA of $u^*$ over $\X$. If some outcomes of products $A$ and
$B$ are not contained in $\X$, it is possible that
$
V_{u^*}(A)>V_{u^*}(B)
$
but
$
V_{u_{\X}^*}(A)<V_{u_{\X}^*}(B).
$
We refer to this phenomenon as a \emph{direction error}. In this case,
the finite-dimensional updating model may not contain a piecewise-linear
representation consistent with the observed preference. See more details
in Appendix~\ref{apx:direction_error}.

These two limitations motivate a dynamic refinement of the breakpoint set. At interaction round $k$, after generating the displayed product list
$
\Sl^k=(S_1^k,\ldots,S_n^k)
$ and observing $\C^k$,
we update the set of breakpoints as:
\begin{equation}
\label{eq:dynamic-breakpoint}
\X^k
:=
\begin{cases}
\displaystyle
\operatorname{sort}\left(
\X^{k-1}
\cup
\bigcup_{i=1}^{n}
\left\{
r_j^{S_i^k}
\mid
j=1,\ldots,n_{S_i^k}
\right\}
\right),
& \text{if } |\C^k|>0, \\[1ex]
\X^{k-1},
& \text{if } |\C^k|=0.
\end{cases}
\end{equation}
which augments the previous breakpoint set with all possible outcomes of the displayed products whenever at least one product is clicked.

For notational convenience, denote the number of breakpoints in $\X^k$ by $N_k+1$, then $\X^k$ can be also represented as
$$
\X^k
=
\{x_1^k<\cdots<x_{N_k+1}^k\}.
$$

When new breakpoints lie inside the current domain, the previous nominal utility is unchanged as a function and is simply represented at the newly inserted points. If a new outcome lies outside the current domain, the domain is expanded and the previous nominal utility is extended linearly using its boundary slopes. Specifically, for
$\tilde{u}_{\X^{k-1}}^{k-1}$ defined on $
\X^{k-1}
=
\{x_1^{k-1}<\cdots<x_{N_{k-1}+1}^{k-1}\},
$
its extended form on $\X^k$, denoted by $\tilde{u}_{\X^k}^{k-1}$, is represented by
\begin{equation*}
\tilde{u}_{\X^k}^{k-1}(x)
=
\begin{cases}
\tilde{u}_{\X^{k-1}}^{k-1}(x_1^{k-1})
+
\beta_1^{k-1}(x-x_1^{k-1}),
\qquad x<x_1^{k-1},
\\[1mm]
\tilde{u}_{\X^{k-1}}^{k-1}(x),
\qquad \qquad \qquad \qquad \quad x\in
[x_1^{k-1},x_{N_{k-1}+1}^{k-1}],
\\[1mm]
\tilde{u}_{\X^{k-1}}^{k-1}(x_{N_{k-1}+1}^{k-1})
+
\beta_{N_{k-1}}^{k-1}
(x-x_{N_{k-1}+1}^{k-1}),
\;x>x_{N_{k-1}+1}^{k-1},
\end{cases}
\end{equation*}
where $\beta_1^{k-1}$ and $\beta_{N_{k-1}}^{k-1}$ are the leftmost and rightmost slopes of $\tilde{u}_{\X^{k-1}}^{k-1}$, respectively. Since the endpoint values are not fixed, 
we define $\mathcal{I}^k:=[x_1^k,x_{N_k+1}^k]$ and the endpoint values in round $k$ by
\begin{equation}
\label{eq:inherited_endpoint_values}
a^k
:=
\tilde{u}_{\X^k}^{k-1}(x_1^k),
\qquad
b^k
:=
\tilde{u}_{\X^k}^{k-1}(x_{N_k+1}^k).
\end{equation}

Accordingly, the admissible utility class at interaction $k$ is
\begin{equation*}
\mathcal U_L^k
:=
\left\{
u:\mathcal{I}^k\to\mathbb R
\ \middle|\
\begin{array}{l}
u(x_1^k)=a^k,\quad
u(x_{N_k+1}^k)=b^k,\\
u\text{ is non-decreasing},\\
|u(x)-u(y)|\leq L|x-y|,
\; \forall x,y\in\mathcal{I}^k
\end{array}
\right\}.
\end{equation*}

\begin{equation*}
\mathcal U_{\X^k}^k
:=
\left\{
u_{\X^k}\in\mathcal U_L^k
\ \middle|\
u_{\X^k}
\text{ is affine on each }
[x_i^k,x_{i+1}^k],
\ i=1,\ldots,N_k
\right\}.
\end{equation*}
And the utility updating model is 
\begin{subequations}\label{eq:updating_PLA_flexible}
\begin{align}
\tilde{u}_{\X^k}^{k}
\in
 \mathop{\arg\min}\limits_{u_{\X^k}
\in\mathcal U_{\X^k}^k}
\quad&\dd_K
\left(
u_{\X^k},
\tilde{u}_{\X^k}^{k-1}
\right)\\
\text{s.t.}
\quad&\E[u_{\X^k}(A)]\geq\E[u_{\X^k}(B)],
\;
\forall (A,B)\in\Q^k.
\end{align}
\end{subequations}

The dynamic breakpoint construction eliminates the direction error for comparisons generated in the current interaction. Indeed, every possible outcome of every product in $\Sl^k$ belongs to $\X^k$. And $
(\Pi_{\X^k}u)(x^k_i)=u(x_i^k)
$ holds for $\forall u\in\mathcal U_{L}^k$. Hence,
$$
V_u(A)\geq V_u(B)
\;\Longleftrightarrow\;
V_{\Pi_{\X^k}u}(A)
\geq
V_{\Pi_{\X^k}u}(B),\quad \forall (A,B)\in\Q^k.
$$

Adding dynamic breakpoint refinement to Algorithm~\ref{alg:rec-alg-fixed} yields \emph{IncVNM}. Algorithm~\ref{alg:incvnm} presents IncVNM without response errors.
\begin{algorithm}
\caption{IncVNM with dynamic breakpoints}
\label{alg:incvnm}
\begin{algorithmic}[1]
    \State \textbf{Input:} Initial breakpoint set $\X^0$, initial utility $\tilde u^0_{\X^0}$ and number of interactions $K$.
    \For{$k=1,\ldots,K$}
        \State Generate $\Sl^k$ from $\tilde u_{\X^{k-1}}^{k-1}$ using Algorithm~\ref{alg:Pro-list-gen}. Observe $\C^k$.
         \If{$|\C^k|=0$}
            \State Set $\Q^k=\varnothing$, $\X^k=\X^{k-1}$ and
            $\tilde{u}_{\X^k}^{k}=\tilde{u}_{\X^{k-1}}^{k-1}$.
        \Else
            \State Construct $\Q^k=\Q(\Sl^k,\C^k)$. Update $\X^{k-1}$ to $\X^k$ by \eqref{eq:dynamic-breakpoint}.
        \State Update the utility using $\tilde u_{\X^{k}}^{k-1}$ (the representation of $\tilde u_{\X^{k-1}}^{k-1}$ on $\X^k$) by solving \eqref{eq:updating_PLA_flexible}, then obtain $\tilde u_{\X^k}^{k}$.
        \EndIf
    \EndFor
    \State \textbf{Output:} $\tilde u_{\X^K}^{K}$.
\end{algorithmic}
\end{algorithm}

With response errors, we solve \eqref{eq:confidence_weighted_update} over $\mathcal U_{\X^k}^k$ using $\tilde{\Q}^k$ instead of solving \eqref{eq:updating_PLA_flexible} in line 8 of Algorithm \ref{alg:incvnm}.


\subsection{Catalog-independent scalability through finite-precision utility representation}
\label{subsec:scalability}
Many recommender systems learn product-specific scores, embeddings, or identifiers \citep{he2020lightgcn,wu2021self,mao2021ultragcn,rajput2023recommender}. Their preference representations therefore grow with the catalog, and new products require additional representations or interaction data. In contrast, IncVNM maintains a single utility function over the common outcome space, enabling direct evaluation of new products from their outcome distributions without product-specific parameters.

For each user, IncVNM stores only a piecewise-linear utility function, whose dimension is determined by the number of breakpoints. Breakpoint growth is therefore the main scalability concern. We control it through a finite-precision representation that uniformly bounds the number of distinct breakpoints.

In practice, numerical outcomes are displayed with finite precision, and differences below this precision are often difficult for users to distinguish and have limited effects on preference learning and product evaluation. For example, a return-rate difference of $0.001$ may not materially affect an investor's assessment. Finite-precision representation also improves computational efficiency by limiting the number of distinct outcomes, breakpoint growth, and the dimension of the utility-updating problem. We therefore retain product outcomes at a prescribed precision $\varepsilon_{\mathrm{prec}}>0$ throughout utility learning.
Specifically, for each outcome $r_i^A$, define its finite-precision representation by
\begin{equation}
\label{eq:precision-rounding}
\bar r_i^A
:=
\varepsilon_{\mathrm{prec}}
\left\lfloor
\frac{r_i^A}{\varepsilon_{\mathrm{prec}}}
+\frac12
\right\rfloor,
\qquad
i=1,\ldots,n_A.
\end{equation}
In other words, $\bar r_i^A$ is obtained by rounding $r_i^A$ to the nearest multiple of $\varepsilon_{\mathrm{prec}}$.
Let $\bar A$ denote the product with outcomes $\bar r_i^A$ and probability vector $\bm p^A$. Then each product $A$ is evaluated through its precision-adjusted counterpart $\bar A$. With a finite-precision representation, we next establish the catalog-independent scalability of the per-user preference representation and utility-updating problem.

\begin{proposition}[Utility error under finite-precision representation]
\label{prop:precision-error}
For any product $A\in\cD$ and any utility function $u\in\U_L$,
$
\left|
V_u(A)-V_u(\bar A)
\right|
\leq
{L\varepsilon_{\mathrm{prec}}}/{2}.
$
\end{proposition}

 Proposition~\ref{prop:precision-error} shows that finite-precision representation introduces a uniformly controlled perturbation in product evaluation.

Accordingly, the dynamic breakpoint set is updated using the precision-adjusted outcomes:
\begin{equation}
\label{eq:dynamic_breakpoint_precision}
\X^k
:=
\begin{cases}
\displaystyle
\operatorname{sort}\left(
\X^{k-1}
\cup
\bigcup_{i=1}^{n}
\left\{
\bar r_j^{S_i^k}
\mid
j=1,\ldots,n_{S_i^k}
\right\}
\right),
& \text{if }|\C^k|>0,\\[1ex]
\X^{k-1},
& \text{if }|\C^k|=0.
\end{cases}
\end{equation}
The following proposition shows that the preference representation and the dimension of each utility update are only determined by the range and precision of the outcome space.
\begin{proposition}[Catalog-independent and history-uniform utility updating]
\label{prop:scalable-update}
Suppose that  precision-adjusted outcomes of all products in $\cD$ lie in a bounded interval
$[\underline{x},\overline{x}]$.
Then
\begin{equation}\label{eq:bp_set_bound}
|\X^k|
\leq \left\lfloor
\frac{\underline{x}-\overline{x}}
{\varepsilon_{\mathrm{prec}}}
\right\rfloor
+|\X^0|+1=:M_{\varepsilon_{\mathrm{prec}}},
\quad\text{for}\;
k=1,2,\ldots
\end{equation}
The numbers of variables and pairwise comparison constraints in the utility update model~\eqref{eq:updating_PLA_flexible} at any interaction round are both of order
$
\mathcal O(M_{\varepsilon_{\mathrm{prec}}}).
$
Consequently, the dimension of per-user utility update is independent of the product-catalog size $|\cD|$ and remains uniformly bounded over all interaction rounds.
\end{proposition}
Hence, IncVNM stores and updates only a bounded finite dimensional utility representation for each user, rather than maintaining product-specific preference parameters. Moreover, the entropy-regularized product-list generation problem admits the closed-form solution in \eqref{eq:softmax} and therefore requires no additional optimization procedure. Consequently, IncVNM achieves compact per-user storage, lightweight online computation, and immediate generalization to newly introduced products and dynamically changing outcome distributions. Proofs of Proposition~\ref{prop:precision-error} and \ref{prop:scalable-update} are in Appendix~\ref{apx:proofs}.

\section{Numerical experiments}
\label{sec:numerical}

We illustrate the performance of IncVNM in a stock recommendation setting.

\tightparagraph{User with response errors}
We consider a simulated risk-averse user whose click behavior is generated based on $V_{u^*}$, where
\begin{equation}
\label{eq:true-utility}
u^*(x)
=
1-0.8\exp(-6x).
\end{equation}
At each interaction, the user clicks the displayed stocks whose oracle expected utilities exceed the median oracle expected utility over the current product catalog.

To model \emph{response errors}, the resulting click sequence is perturbed with probability $\varepsilon_{\mathrm{noise}}$. Conditional on a perturbation, one feasible operation is selected at random and applied to the click sequence: swapping two clicked stocks, inserting an unclicked stock, or removing a clicked stock. These operations mimic local ranking errors, accidental clicks, and overlooked products, respectively.

\tightparagraph{Rolling-window protocol}
We collect monthly returns from January 2022 to April 2026 for 463 constituent stocks of the S\&P 500 with complete data over the sample period.\footnote{Data downloaded from \url{https://finance.yahoo.com/}.} We conduct a rolling-window experiment using a 24-month historical window. The first window covers January 2022 to December 2023.
Within each window, we simulate 20 interactions, each displaying 8 stocks.

After these 20 interactions, the historical window is shifted forward
by one month. Thus, the second window covers February 2022 to January
2024.
This procedure continues until the final window, from May 2024 to April 2026. The experiment therefore contains 29 rolling windows and 580 interactions in total. Within each window, the empirical outcome distribution of each stock remains unchanged and is re-estimated when the window moves forward.

\begin{figure}[htbp]
    \centering
    \includegraphics[width=0.6\linewidth]{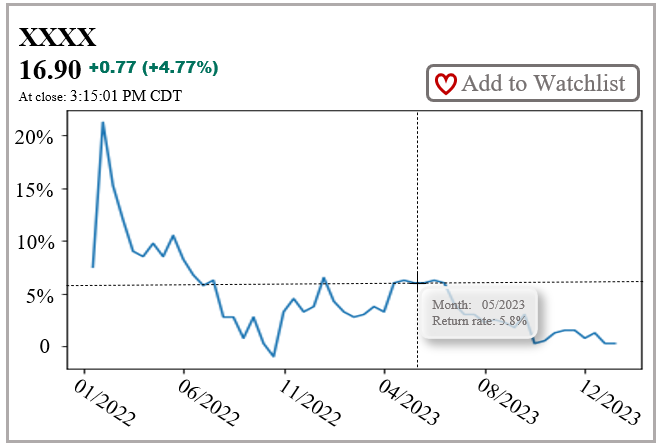}
    \caption{Historical information displayed on the webpage.}
    \Description{A webpage displaying historical monthly returns and summary information for several stocks.}
    \label{fig:stock-web}
\end{figure}

\tightparagraph{Empirical return distributions}
Each displayed stock is shown together with its 24-month historical returns. We round each monthly return to the nearest multiple of $\varepsilon_{\mathrm{prec}}=0.01$.
For notational simplicity, we henceforth use $A$ and $r_i^A$ to
denote their precision-adjusted counterparts $\bar A$ and
$\bar r_i^A$, respectively. Figure~\ref{fig:stock-web} provides an example.

At interaction $k$, the displayed product list is
$
\Sl^k
=
(S_1^k,\ldots,S_8^k)
\subseteq
\cD.
$
For each displayed stock $S_i^k$, let
$
\bm r^{S_i^k}
=
(
r_1^{S_i^k},
\ldots,
r_{24}^{S_i^k}
)
$
denote its monthly return rates in the current 24-month historical window. The VNM expected utility of $S_i^k$ under a utility function $u\in\U_L$ is estimated by the sample mean:
\begin{equation}
\label{eq:V_k}
V_u^k(S_i^k)
:=
\frac{1}{24}
\sum_{j=1}^{24}
u(r_j^{S_i^k}),
\qquad
i=1,\ldots,8.
\end{equation}

 \subsection{Baseline methods}
\label{sec:baseline}

We compare IncVNM with four representative methods covering three
recommendation paradigms. The first category consists of
\emph{outcome-based scoring methods}, including expected return
(ER) and mean--variance (MV) \citep{markowitz1952portfolio}, which rank
stocks using fixed statistics of their current return distributions
without learning from click feedback. The second category is
\emph{product-level ranking methods}, represented by discounted upper
confidence bound (DUCB)
\citep{auer2002finite,garivier2011upper}, which learns discounted
product-level rewards solely from click feedback without accounting for
the products' outcome distributions. The third category is
\emph{distribution-level ranking methods}, represented by distribution-aware Bayesian personalized
ranking (DistBPR), our
distribution-level extension of BPR
\citep{rendle2009bpr}. DistBPR learns a ranking score
from pairwise click feedback while representing each stock through its outcome distribution. The implementation details of these methods can be found
in Appendix~\ref{app:baseline_details}.

\subsection{Performance evaluation metrics}
\label{sec:metrics}

Let
$
\mathcal M
:=
\{\mathrm{IncVNM},\mathrm{ER},\mathrm{MV},
\mathrm{DistBPR},\mathrm{DUCB}\}
$
denote the set of evaluated methods. At interaction $k$, each method
$m\in\mathcal M$ assigns a score $s_m^k(A)$ to every stock
$A\in\cD$. In particular,
$$
s_*^k(A)
:=
V_{u^*}^k(A),
\qquad
s_{\mathrm{IncVNM}}^k(A)
:=
V_{\tilde u_{\X^{k-1}}^{k-1}}^k(A),
$$
where $s_*^k$ denotes the oracle score.

For any scoring function $s:\cD\to\R$, let
\begin{equation}
\mathcal O(s)
:=
\left(
O_1(s),\ldots,O_{|\cD|}(s)
\right)
\label{eq:ranking-list}
\end{equation}
denote the ordered list of stocks sorted in descending order of
$s(\cdot)$.

For any ordered list $\mathcal O$, define
\begin{equation}
\operatorname{rank}(A;\mathcal O)
:=
i
\quad
\text{if }
A
\text{ is the }i\text{-th stock in }\mathcal O.
\label{eq:rank-definition}
\end{equation}

We use two widely used metrics to evaluate utility-learning and
recommendation performance.

\emph{\underline{Ranking difference (RD)}}.
For an ordered list
$
\mathcal O=(O_1,\ldots,O_{\ell}),
$
the ranking difference at interaction $k$ is defined as
\begin{equation}
\operatorname{RD}^k(\mathcal O)
:=
\sum_{i=1}^{\ell}
\left|
i
-
\operatorname{rank}
\left(
O_i;\mathcal O(s_*^k)
\right)
\right|.
\label{eq:RD}
\end{equation}
This metric measures the discrepancy between the positions of the
stocks in $\mathcal O$ and their positions in the oracle ranking. A
smaller value indicates greater ranking accuracy. It is closely
related to Spearman's footrule distance
\citep{diaconis1977spearman}.

\emph{\underline{Normalized half-life utility (nHLU)}}.
To account for position dependent user attention, we adapt the
half-life utility metric
\citep{breese1998empirical}. Define the oracle relevance of stock
$A$ at interaction $k$ as
\begin{equation}
\operatorname{rel}^k(A)
:=
\frac{
s_*^k(A)
-
\min_{B\in\cD}s_*^k(B)
}{
\max_{B\in\cD}s_*^k(B)
-
\min_{B\in\cD}s_*^k(B)
+
\epsilon
},
\label{eq:oracle-relevance}
\end{equation}
where $\epsilon>0$ is a small constant used to avoid division by
zero.
For position $i$, define the viewing weight
$
w_i
:=
2^{-\frac{i-1}{h-1}},
$
where $h>1$ is the half-life parameter. We set $h=4$, so that the
fourth position receives one half of the weight assigned to the first
position. For an ordered list
$
\mathcal O=(O_1,\ldots,O_\ell),
$
its normalized half-life utility is
\begin{equation*}
\operatorname{nHLU}^k(\mathcal O)
:=
{
\left(\displaystyle
\sum_{i=1}^{\ell}
w_i\operatorname{rel}^k(O_i)
\right)}\bigg/\left({
\displaystyle
\sum_{i=1}^{\ell}
w_i\operatorname{rel}^k
\left(
O_i(s_*^k)
\right)
+
\epsilon
}\right).
\end{equation*}
A larger value, approaching 1, indicates that stocks with high oracle relevance are
placed closer to the top of $\mathcal O$.

\tightparagraph{Top-8 ranking evaluation metrics}
To evaluate utility-learning performance, we set $\mathcal O$ to the
top-$8$ list induced by the score of method $m$, which is 
$
\mathcal O_{@8}^k(m)
:=
\left(
O_1(s_m^k),\ldots,O_8(s_m^k)
\right).
$
Define
$$
\operatorname{RD}_{@8}^k(m)
:=
\operatorname{RD}^k
\left(
\mathcal O_{@8}^k(m)
\right),
\quad
\operatorname{nHLU}_{@8}^k(m)
:=
\operatorname{nHLU}^k
\left(
\mathcal O_{@8}^k(m)
\right),
$$
which evaluate the performance of scoring or utility learning.
\tightparagraph{Recommendation evaluation metrics}
Let
$
\Sl_m^k
:=
\left(
S_{m,1}^k,\ldots,S_{m,8}^k
\right)
$
denote the product list actually generated by method $m$ at
interaction $k$. We define
$$
\operatorname{RD}_{\mathrm{pl}}^k(m)
:=
\operatorname{RD}^k(\Sl_m^k),
\qquad
\operatorname{nHLU}_{\mathrm{pl}}^k(m)
:=
\operatorname{nHLU}^k(\Sl_m^k),
$$
which evaluate
each method's
recommendation policy quality.

\subsection{Results with response errors}
\label{sec:results-error}

\begin{figure}[t]
    \centering
    \includegraphics[width=0.7\linewidth]{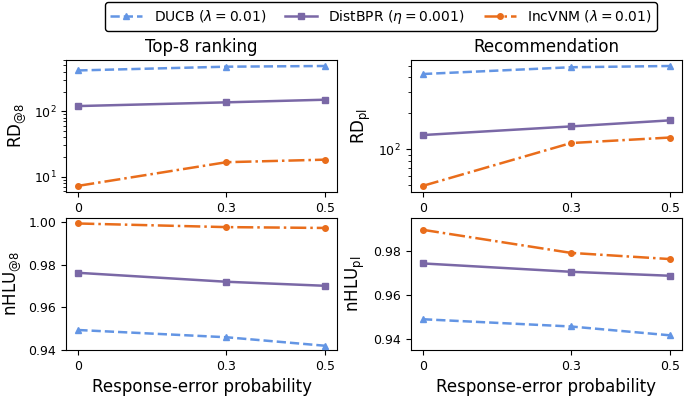}
    \caption{Performance under different
response-error probabilities. The RD metrics are displayed
on a logarithmic scale.}
    \Description{Four plots compare the cumulative ranking-difference and normalized half-life-utility metrics of the evaluated methods under different response-error probabilities.}
    \label{fig:error_comparison}
\end{figure}

We evaluate ER, MV, DistBPR with $\eta=0.001$, DUCB with
$\lambda=0.01$, and IncVNM with $\lambda=0.01$ under
$\varepsilon_{\mathrm{noise}}\in\{0,0.3,0.5\}$.
Table~\ref{tab:error_comparison} reports the cumulative-average utility-learning and recommendation performance after 200, 400, and 580 interactions. Each entry is the mean of the corresponding cumulative-average metric over 15 runs, comprising five repetitions for each of three random seeds. Figure~\ref{fig:error_comparison} presents the metrics of methods after 580 interactions under different error rates.
The results lead to the following observations.

\begin{table}[!htbp]
\centering
\caption{Cumulative-average ranking performance under different response-error probabilities.}
\label{tab:error_comparison}

\fontsize{8pt}{9.6pt}\selectfont
\setlength{\tabcolsep}{0.8pt}
\renewcommand{\arraystretch}{0.95}

\newcommand{\errhead}[1]{{\fontsize{7.5pt}{9pt}\selectfont #1}}

\begin{tabularx}{\textwidth}{
@{}
>{\centering\arraybackslash}p{1.40cm}
>{\centering\arraybackslash}p{1.40cm}
>{\centering\arraybackslash}p{0.50cm}
*{2}{>{\centering\arraybackslash}p{0.95cm}}
*{9}{>{\centering\arraybackslash}X}
@{}
}
\toprule
\multicolumn{5}{c}{}
& \multicolumn{3}{c}{$\varepsilon_{\mathrm{noise}}=0$}
& \multicolumn{3}{c}{$\varepsilon_{\mathrm{noise}}=0.3$}
& \multicolumn{3}{c}{$\varepsilon_{\mathrm{noise}}=0.5$} \\
\cmidrule(lr){6-8}
\cmidrule(lr){9-11}
\cmidrule(lr){12-14}
\multicolumn{3}{c}{}
& \errhead{ER} & \errhead{MV}
& \errhead{DistBPR} & \errhead{DUCB} & \errhead{IncVNM}
& \errhead{DistBPR} & \errhead{DUCB} & \errhead{IncVNM}
& \errhead{DistBPR} & \errhead{DUCB} & \errhead{IncVNM} \\
\midrule

\multirow{6}{1.40cm}{\centering\shortstack{Top-8\\ranking}}
& \multirow{3}{*}{$\mathrm{RD}_{@8}\downarrow$}
& 200 & 567.90 & 170.30 & 155.22 & 729.76 & \textbf{15.63} & 165.69 & 761.91 & \textbf{28.33} & 171.36 & 746.47 & \textbf{26.17} \\
& & 400 & 409.10 & 107.35 & 122.69 & 467.52 & \textbf{8.90} & 148.82 & 509.42 & \textbf{19.20} & 162.80 & 524.82 & \textbf{18.67} \\
& & 580 & 378.86 & 117.48 & 119.92 & 419.64 & \textbf{7.26} & 136.71 & 478.84 & \textbf{16.66} & 149.91 & 490.61 & \textbf{18.22} \\

\addlinespace[1pt]
\cmidrule(lr){2-14}
\addlinespace[1pt]

& \multirow{3}{*}{$\mathrm{nHLU}_{@8}\uparrow$}
& 200 & 0.9192 & 0.9736 & 0.9693 & 0.9168 & \textbf{0.9988} & 0.9653 & 0.9163 & \textbf{0.9978} & 0.9629 & 0.9153 & \textbf{0.9975} \\
& & 400 & 0.9433 & 0.9799 & 0.9727 & 0.9371 & \textbf{0.9993} & 0.9678 & 0.9361 & \textbf{0.9973} & 0.9642 & 0.9314 & \textbf{0.9974} \\
& & 580 & 0.9491 & 0.9792 & 0.9763 & 0.9494 & \textbf{0.9995} & 0.9721 & 0.9460 & \textbf{0.9978} & 0.9702 & 0.9420 & \textbf{0.9974} \\

\midrule

\multirow{6}{1.40cm}{\centering\shortstack{Recom-\\mendation}}
& \multirow{3}{*}{$\mathrm{RD}_{\mathrm{pl}}\downarrow$}
& 200 & 567.90 & 170.30 & 158.74 & 745.27 & \textbf{50.61} & 177.77 & 777.42 & \textbf{65.50} & 186.57 & 761.98 & \textbf{77.83} \\
& & 400 & 409.10 & 107.35 & 132.94 & 475.27 & \textbf{49.61} & 171.63 & 517.17 & \textbf{95.97} & 188.23 & 532.58 & \textbf{92.49} \\
& & 580 & 378.86 & 117.48 & 131.16 & 424.99 & \textbf{49.31} & 154.87 & 484.19 & \textbf{112.43} & 173.96 & 495.96 & \textbf{124.98} \\

\addlinespace[1pt]
\cmidrule(lr){2-14}
\addlinespace[1pt]

& \multirow{3}{*}{$\mathrm{nHLU}_{\mathrm{pl}}\uparrow$}
& 200 & 0.9192 & 0.9736 & 0.9667 & 0.9158 & \textbf{0.9889} & 0.9636 & 0.9152 & \textbf{0.9875} & 0.9618 & 0.9143 & \textbf{0.9840} \\
& & 400 & 0.9433 & 0.9799 & 0.9703 & 0.9366 & \textbf{0.9884} & 0.9658 & 0.9356 & \textbf{0.9779} & 0.9626 & 0.9309 & \textbf{0.9779} \\
& & 580 & 0.9491 & 0.9792 & 0.9744 & 0.9490 & \textbf{0.9897} & 0.9706 & 0.9457 & \textbf{0.9792} & 0.9688 & 0.9417 & \textbf{0.9764} \\
\bottomrule
\end{tabularx}
\end{table}
\noindent\textbf{(a) IncVNM performs best and 
most robust.}
IncVNM consistently achieves the best utility-learning and recommendation performance across all response-error probabilities. 

\noindent\textbf{(b) Random-outcome recommendation requires both distribution level representation and risk-preference modeling.}
IncVNM and DistBPR outperform DUCB because they represent each stock through its current return distribution, whereas DUCB learns product-level rewards that transfer poorly across rolling windows. Risk modeling is also important: MV outperforms ER by partially capturing risk aversion through variance of stocks, while IncVNM further learns the user's risk attitude through a VNM utility function. Therefore, effective recommendation for random-outcome products requires both distribution-level representation and explicit risk-preference modeling.

\noindent\textbf{(c) Response errors degrade all feedback-driven methods.}
ER and MV are unaffected because they do not learn from clicks. DistBPR and DUCB only passively fit erroneous feedback, whereas IncVNM limits its influence through confidence-weighted relaxation and hence consistently retains superior performance.

We also evaluate the methods under different parameter settings in the error-free case. The results are reported in Appendix~\ref{sec:results-errorfree}.

\subsection{Scalability analysis}
\label{sec:scalability-experiment}

\begin{figure}[htbp]
    \centering
    \includegraphics[width=0.7\linewidth]{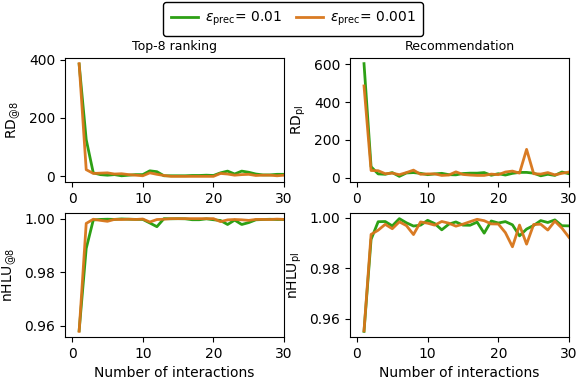}
    \caption{Performance of IncVNM under different precisions.}
    \Description{Four plots compare utility-learning and recommendation metrics over 30 interactions under two finite-precision levels.}
    \label{fig:precision-performance}
\end{figure}

We numerically examine the scalability properties established in
Section~\ref{subsec:scalability}.
We test IncVNM with
$\varepsilon_{\mathrm{prec}}=0.01$ and $0.001$ for 30 interactions. All other settings remain unchanged from the previous sections.
Figure~\ref{fig:precision-performance} reports the performance metrics,
while Table~\ref{tab:precision-scalability} reports the initial and final
breakpoint numbers and the CPU time\footnote{CPU time is averaged over five runs with a 12th Gen Intel(R) Core(TM) i5-12490F.}. 

\begin{table}[htbp]
    \centering
    \caption{Breakpoints and CPU time by precision.}
    \label{tab:precision-scalability}
    \small
    \renewcommand{\arraystretch}{0.95}
    \begin{tabular}{cccc}
        \toprule
        $\varepsilon_{\mathrm{prec}}$
        & $|\X^0|$
        & $|\X^{30}|$
        & CPU time (s) \\
        \midrule
        $0.01$  & 2 & 53  & 42.81  \\
        $0.001$ & 2 & 213 & 462.06 \\
        \bottomrule
    \end{tabular}
\end{table}
The results show that the finer precision level, $\varepsilon_{\mathrm{prec}}=0.001$, yields
slightly better performance. However, it substantially
increases the number of breakpoints and the CPU time. These results
illustrate the accuracy--computation trade-off and are consistent with
Proposition~\ref{prop:precision-error}.

\begin{figure}[htbp]
    \centering
    \includegraphics[width=0.6\linewidth]{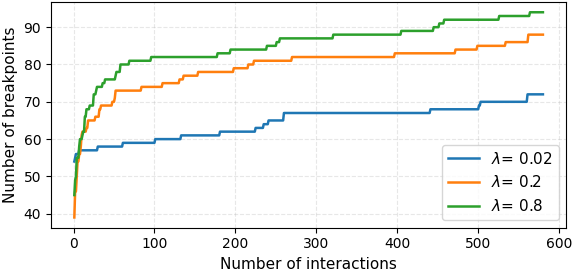}
    \caption{Breakpoint growth under different entropy-regularization parameters.}
    \Description{A line plot shows the number of breakpoints over interactions for several entropy-regularization parameters.}
    \label{fig:breakpoint-growth}
\end{figure}

We further report the breakpoint growth under different
exploration rates $\lambda$, with
$\varepsilon_{\mathrm{prec}}=0.01$, in
Figure~\ref{fig:breakpoint-growth}. A larger $\lambda$ induces broader
exploration and therefore introduces more breakpoints. In particular,
when $\lambda=0.8$, the sampling distribution used for product-list
generation is close to uniform. Nevertheless, the breakpoint growth
slows over time, and the number of breakpoints remains below the upper
bound established in Proposition~\ref{prop:scalable-update},
namely, $M_{\varepsilon_{\mathrm{prec}}}=167$ in this experiment, even
for a large value of $\lambda$.

\section{Conclusion}
\label{sec:conclusion}

This paper proposes IncVNM, an interactive recommender system for products with random outcomes, which addresses two key limitations in existing research on recommendation under outcome uncertainty. First, IncVNM learns an interpretable and transferable user-specific VNM utility function that captures individual risk attitudes and evaluates products through their outcome distributions. It can therefore directly accommodate newly introduced products and time-varying outcome distributions without requiring user-product historical data. Second, IncVNM learns VNM utility incrementally from ordinal click feedback generated during ordinary webpage browsing, thereby integrating preference elicitation and recommendation into an interactive process. Future research may evaluate IncVNM through online experiments with real users.

\bibliographystyle{plainnat}
\bibliography{references}

@inproceedings{mao2021ultragcn,
  title={UltraGCN: ultra simplification of graph convolutional networks for recommendation},
  author={Mao, Kelong and Zhu, Jieming and Xiao, Xi and Lu, Biao and Wang, Zhaowei and He, Xiuqiang},
  booktitle={Proceedings of the 30th ACM international conference on information \& knowledge management},
  pages={1253--1262},
  year={2021}
}

@inproceedings{wu2021self,
  title={Self-supervised graph learning for recommendation},
  author={Wu, Jiancan and Wang, Xiang and Feng, Fuli and He, Xiangnan and Chen, Liang and Lian, Jianxun and Xie, Xing},
  booktitle={Proceedings of the 44th international ACM SIGIR conference on research and development in information retrieval},
  pages={726--735},
  year={2021}
}

@inproceedings{christiano2017deep,
  title={Deep reinforcement learning from human preferences},
  author={Christiano, Paul F and Leike, Jan and Brown, Tom and Martic, Miljan and Legg, Shane and Amodei, Dario},
  booktitle={Proceedings of the 31st International Conference on Neural Information Processing Systems},
  pages={4302--4310},
  year={2017}
}

@article{guo2024utility,
  title={Utility preference robust optimization with moment-type information structure},
  author={Guo, Shaoyan and Xu, Huifu and Zhang, Sainan},
  journal={Operations Research},
  volume={72},
  number={5},
  pages={2241--2261},
  year={2024},
  publisher={INFORMS}
}

@article{hu2015robust,
  title={Robust decision making over a set of random targets or risk-averse utilities with an application to portfolio optimization},
  author={Hu, Jian and Mehrotra, Sanjay},
  journal={IIE Transactions},
  volume={47},
  number={4},
  pages={358--372},
  year={2015},
  publisher={Taylor \& Francis}
}

@article{chen2026eliciting,
  title={Eliciting Von Neumann--Morgenstern utility from discrete choices with response error},
  author={Chen, Bo and Liu, Jia},
  journal={Mathematical Programming},
  pages={1--54},
  year={2026},
  publisher={Springer}
}

@article{hey1994investigating,
  title={Investigating generalizations of expected utility theory using experimental data},
  author={Hey, John D and Orme, Chris},
  journal={Econometrica: Journal of the Econometric Society},
  volume={62},
  number={6},
  pages={1291--1326},
  year={1994},
  publisher={JSTOR}
}

@article{eliashberg1985measurement,
  title={A measurement error approach for modeling consumer risk preference},
  author={Eliashberg, Jehoshua and Hauser, John R},
  journal={Management Science},
  volume={31},
  number={1},
  pages={1--25},
  year={1985},
  publisher={INFORMS}
}

@article{wakker1996eliciting,
  title={Eliciting von Neumann-Morgenstern utilities when probabilities are distorted or unknown},
  author={Wakker, Peter and Deneffe, Daniel},
  journal={Management science},
  volume={42},
  number={8},
  pages={1131--1150},
  year={1996},
  publisher={INFORMS}
}

@article{farquhar1984state,
  title={State of the art—utility assessment methods},
  author={Farquhar, Peter H},
  journal={Management science},
  volume={30},
  number={11},
  pages={1283--1300},
  year={1984},
  publisher={INFORMS}
}

@article{zhang2025modified,
  title={Modified Polyhedral Method for Elicitation of Shape-Free Utility and Conservatism Reduction in Robust Optimization},
  author={Zhang, Sainan and Guo, Shaoyan and Sim, Melvyn and Xu, Huifu},
  journal={arXiv preprint arXiv:2503.23269},
  year={2025}
}

@inproceedings{breese1998empirical,
  title={Empirical analysis of predictive algorithms for collaborative filtering},
  author={Breese, John S and Heckerman, David and Kadie, Carl},
  booktitle={Proceedings of the 14th Conference on Uncertainty in Artificial Intelligence},
  pages={43--52},
  year={1998}
}

@article{diaconis1977spearman,
  title={Spearman's footrule as a measure of disarray},
  author={Diaconis, Persi and Graham, Ronald L},
  journal={Journal of the Royal Statistical Society Series B: Statistical Methodology},
  volume={39},
  number={2},
  pages={262--268},
  year={1977},
  publisher={Oxford University Press}
}

@inproceedings{garivier2011upper,
  title={On upper-confidence bound policies for switching bandit problems},
  author={Garivier, Aur{\'e}lien and Moulines, Eric},
  booktitle={International conference on algorithmic learning theory},
  pages={174--188},
  year={2011},
  publisher={Springer}
}

@article{rajput2023recommender,
  title={Recommender systems with generative retrieval},
  author={Rajput, Shashank and Mehta, Nikhil and Singh, Anima and Hulikal Keshavan, Raghunandan and Vu, Trung and Heldt, Lukasz and Hong, Lichan and Tay, Yi and Tran, Vinh and Samost, Jonah and others},
  journal={Advances in Neural Information Processing Systems},
  volume={36},
  pages={10299--10315},
  year={2023}
}

@inproceedings{he2020lightgcn,
  title={Lightgcn: Simplifying and powering graph convolution network for recommendation},
  author={He, Xiangnan and Deng, Kuan and Wang, Xiang and Li, Yan and Zhang, Yongdong and Wang, Meng},
  booktitle={Proceedings of the 43rd International ACM SIGIR conference on research and development in Information Retrieval},
  pages={639--648},
  year={2020}
}

@inproceedings{sun2019bert4rec,
  title={BERT4Rec: Sequential recommendation with bidirectional encoder representations from transformer},
  author={Sun, Fei and Liu, Jun and Wu, Jian and Pei, Changhua and Lin, Xiao and Ou, Wenwu and Jiang, Peng},
  booktitle={Proceedings of the 28th ACM international conference on information and knowledge management},
  pages={1441--1450},
  year={2019}
}

@inproceedings{kang2018self,
  title={Self-attentive sequential recommendation},
  author={Kang, Wang-Cheng and McAuley, Julian},
  booktitle={2018 IEEE international conference on data mining (ICDM)},
  pages={197--206},
  year={2018},
  organization={IEEE}
}

@inproceedings{hidasi2016parallel,
  title={Parallel recurrent neural network architectures for feature-rich session-based recommendations},
  author={Hidasi, Bal{\'a}zs and Quadrana, Massimo and Karatzoglou, Alexandros and Tikk, Domonkos},
  booktitle={Proceedings of the 10th ACM conference on recommender systems},
  pages={241--248},
  year={2016}
}

@article{linden2003amazon,
  title={Amazon. com recommendations: Item-to-item collaborative filtering},
  author={Linden, Greg and Smith, Brent and York, Jeremy},
  journal={IEEE Internet computing},
  volume={7},
  number={1},
  pages={76--80},
  year={2003},
  publisher={IEEE}
}

@incollection{pazzani2007content,
  title={Content-based recommendation systems},
  author={Pazzani, Michael J and Billsus, Daniel},
  booktitle={The adaptive web: methods and strategies of web personalization},
  pages={325--341},
  year={2007},
  publisher={Springer}
}

@article{adomavicius2005toward,
  title={Toward the next generation of recommender systems: A survey of the state-of-the-art and possible extensions},
  author={Adomavicius, Gediminas and Tuzhilin, Alexander},
  journal={IEEE transactions on knowledge and data engineering},
  volume={17},
  number={6},
  pages={734--749},
  year={2005},
  publisher={IEEE}
}

@inproceedings{resnick1994grouplens,
  title={Grouplens: An open architecture for collaborative filtering of netnews},
  author={Resnick, Paul and Iacovou, Neophytos and Suchak, Mitesh and Bergstrom, Peter and Riedl, John},
  booktitle={Proceedings of the 1994 ACM conference on Computer supported cooperative work},
  pages={175--186},
  year={1994}
}

@article{wei2026coordinate,
  title={Coordinate-wise Polyhedral Method for Eliciting Multivariate Linear Utility and Univariate Nonlinear Utility Functions},
  author={Wei, Jiaxin and Liu, Jia and Xu, Huifu},
  journal={arXiv preprint arXiv:2606.13418},
  year={2026}
}

@article{markowitz1952portfolio,
  title={Portfolio Selection},
  author={Markowitz, Harry},
  journal={The Journal of Finance},
  volume={7},
  number={1},
  pages={77--91},
  year={1952}
}

@book{von1944theory,
  title={Theory of Games and Economic Behavior},
  author={von Neumann, John and Morgenstern, Oskar},
  publisher={Princeton University Press},
  year={1944}
}

@inproceedings{rendle2009bpr,
  title={BPR: Bayesian Personalized Ranking from Implicit Feedback},
  author={Rendle, Steffen and Freudenthaler, Christoph and Gantner, Zeno and Schmidt-Thieme, Lars},
  booktitle={Proceedings of the 25th Conference on Uncertainty in Artificial Intelligence},
  pages={452--461},
  year={2009}
}

@article{koren2009matrix,
  title={Matrix Factorization Techniques for Recommender Systems},
  author={Koren, Yehuda and Bell, Robert and Volinsky, Chris},
  journal={Computer},
  volume={42},
  number={8},
  pages={30--37},
  year={2009}
}

@article{auer2002finite,
  title={Finite-time Analysis of the Multiarmed Bandit Problem},
  author={Auer, Peter and Cesa-Bianchi, Nicol\`o and Fischer, Paul},
  journal={Machine Learning},
  volume={47},
  number={2},
  pages={235--256},
  year={2002}
}

@inproceedings{li2010contextual,
  title={A Contextual-Bandit Approach to Personalized News Article Recommendation},
  author={Li, Lihong and Chu, Wei and Langford, John and Schapire, Robert E.},
  booktitle={Proceedings of the 19th International Conference on World Wide Web},
  pages={661--670},
  year={2010}
}

@article{armbruster2015decision,
  title={Decision Making Under Uncertainty When Preference Information Is Incomplete},
  author={Armbruster, Benjamin and Delage, Erick},
  journal={Management Science},
  volume={61},
  number={1},
  pages={111--128},
  year={2015}
}

\appendix
\section{Kantorovich distance}
\label{apx:kanto}

We first provide the definition and an equivalent representation of the Kantorovich distance used in the utility-updating models.

\begin{definition}[Kantorovich distance]
Define the class of $1$-Lipschitz test functions by
$$
\mathcal G_1
:=
\left\{
g:[\underline{x},\overline{x}]\to\R
\ \middle|\
|g(x)-g(y)|
\leq
|x-y|,
\quad
\forall x,y\in[\underline{x},\overline{x}]
\right\}.
$$
Let
$$
\begin{aligned}
   {\U}:=\Big\{u:[\underline{x},\overline{x}]&\rightarrow[a,b] \mid u(\underline{x})=a, \ u(\overline{x})=b,\\
&u \text{ is nondecreasing
and right-continuous}
\Big\}. 
\end{aligned}
$$
For any $u,v\in\U$, the Kantorovich distance between them is defined as
\begin{equation}
\label{eq:uv-Kant-dist}
\dd_K(u,v)
:=
\sup_{g\in\mathcal G_1}
\left\{
\int_{\underline{x}}^{\overline{x}}g(x)\,du(x)
-
\int_{\underline{x}}^{\overline{x}}g(x)\,dv(x)
\right\},
\end{equation}
where the integrals are the Lebesgue--Stieltjes integrals.
\end{definition}

Then we show that the Kantorovich distance used in this paper coincides with the $L^1$ distance between two normalized utility functions. Since $\mathcal G_1$ is symmetric, namely, $g\in\mathcal G_1$ implies $-g\in\mathcal G_1$, definition \eqref{eq:uv-Kant-dist} can equivalently be written as
$$
\dd_K(u,v)
=
\sup_{g\in\mathcal G_1}
\left|
\int_{\underline{x}}^{\overline{x}}g(x)\,du(x)
-
\int_{\underline{x}}^{\overline{x}}g(x)\,dv(x)
\right|.
$$

For any $u,v\in\U$, the endpoint normalization gives
$
u(\underline{x})=v(\underline{x})=a
\;\text{and}\;
u(\overline{x})=v(\overline{x})=b.
$
Since $g\in\mathcal G_1$ are Lipschitz continuous, hence absolutely continuous. Integration by parts yields
\begin{equation}  
\int_{\underline{x}}^{\overline{x}}g(x)\,du(x)
-
\int_{\underline{x}}^{\overline{x}}g(x)\,dv(x)
=
-
\int_{\underline{x}}^{\overline{x}}
g'(x)\bigl(u(x)-v(x)\bigr)\,dx.
\label{eq:kanto-integration-by-parts}
\end{equation}
Because $|g'(x)|\leq 1$ almost everywhere, we have
$$
\dd_K(u,v)
\leq
\int_{\underline{x}}^{\overline{x}}
|u(x)-v(x)|\,dx.
$$

To establish the reverse inequality, define
$$
g^*(x)
:=
-
\int_{\underline{x}}^x
\operatorname{sgn}\bigl(u(t)-v(t)\bigr)\,dt.
$$
Since
$
\left|(g^*)'(x)\right|
=
\left|
\operatorname{sgn}\bigl(u(x)-v(x)\bigr)
\right|
\leq 1
$
almost everywhere, we have $g^*\in\mathcal G_1$. Substituting $g^*$ into \eqref{eq:kanto-integration-by-parts} gives
$$
-
\int_{\underline{x}}^{\overline{x}}
(g^*)'(x)\bigl(u(x)-v(x)\bigr)\,dx
=
\int_{\underline{x}}^{\overline{x}}
|u(x)-v(x)|\,dx.
$$
Consequently,
\begin{equation}
\label{eq:kanto-L1-equivalence}
\dd_K(u,v)
=
\int_{\underline{x}}^{\overline{x}}
|u(x)-v(x)|\,dx,
\qquad
\forall u,v\in\U.
\end{equation}
 Geometrically, the Kantorovich distance is the total area between their graphs. In particular, for two piecewise-linear utility functions, \eqref{eq:kanto-L1-equivalence} can be evaluated by summing the areas between their linear segments.

\section{Reformulation of the utility updating problem}
\label{apx:reform_updating}

Fix an interaction round $k$ and a current breakpoint set
$$
\X
=
\{x_1<\cdots<x_{N+1}\}.
$$
$\tilde{u}_{\X}^{k-1}$ denotes the nominal utility after interaction $k-1$ on the breakpoint set $\X$. 

Recall that the piecewise-linear utility-updating problem is
\begin{subequations}
\label{eq:app-updating-PLA}
\begin{align}
\tilde{u}_{\X}^{k}
\in
\arg\min_{u_{\X}\in\U_{\X}}
\quad
&
\dd_K
\left(
u_{\X},
\tilde{u}_{\X}^{k-1}
\right)
\label{eq:app-updating-PLA-obj}
\\
\text{s.t.}\quad
&
V_{u_{\X}}(A)
\geq
V_{u_{\X}}(B),
\qquad
\forall (A,B)\in\Q^k.
\label{eq:app-updating-PLA-cons}
\end{align}
\end{subequations}

For any $u_{\X}\in\U_{\X}$, define
\begin{align*}
   & \alpha_j
:=
u_{\X}(x_j),
\qquad
j=1,\ldots,N+1,\\
\text{and}\quad
&\beta_j
:=
\frac{\alpha_{j+1}-\alpha_j}
{x_{j+1}-x_j},
\qquad
j=1,\ldots,N.
\end{align*}
Then
$
u_{\X}(x)
=
\alpha_j+\beta_j(x-x_j),
\quad
x\in[x_j,x_{j+1}],
\quad
j=1,\ldots,N.$

Similarly, for $\tilde{u}_{\X}^{k-1}$, define
\begin{align*}
  &\tilde{\alpha}_j^{k-1}
:=
\tilde{u}_{\X}^{k-1}(x_j),\;
j=1,\ldots,N+1,\\
\text{and}\quad
&\tilde{\beta}_j^{k-1}
:=
\frac{
\tilde{\alpha}_{j+1}^{k-1}
-
\tilde{\alpha}_{j}^{k-1}
}{
x_{j+1}-x_j
},
\;
j=1,\ldots,N.  
\end{align*}

For any test function $g\in\mathcal G_1$ defined in Appendix~\ref{apx:kanto}, the Lebesgue--Stieltjes integral over a piece-wise linear utility function satisfies
$$
\int_{x_1}^{x_{N+1}}
g(z)\,du_{\X}(z)
=
\sum_{j=1}^{N}
\beta_j
\int_{x_j}^{x_{j+1}}g(t)\,dt.
$$
Therefore,
\begin{equation}
\label{eq:kanto-slope-form}
\dd_K
\left(
u_{\X},
\tilde{u}_{\X}^{k-1}
\right)
=
\sup_{g\in\mathcal G_1}
\sum_{j=1}^{N}
\left(
\beta_j-\tilde{\beta}_j^{k-1}
\right)
\int_{x_j}^{x_{j+1}}g(t)\,dt.
\end{equation}

The endpoint constraints ensure that
$$
u_{\X}(x_1)
=
\tilde{u}_{\X}^{k-1}(x_1)
\quad\text{and}\quad
u_{\X}(x_{N+1})
=
\tilde{u}_{\X}^{k-1}(x_{N+1}).
$$
Consequently,
$
\sum_{j=1}^{N}
\left(
\beta_j-\tilde{\beta}_j^{k-1}
\right)
(x_{j+1}-x_j)
=
0.
$
Hence, adding a constant to $g$ does not change the objective in \eqref{eq:kanto-slope-form}, and we may impose
$
g(x_1)=0
$
without loss of generality.

Define
$$
h_j
:=
x_{j+1}-x_j,
\quad
\chi_j
:=
g(x_{j+1})-g(x_j),
\quad
j=1,\ldots,N.
$$
It follows that
$$
g(x_j)
=
\sum_{i=1}^{j-1}\chi_i
\quad\text{and}\quad
g(x_{j+1})+g(x_j)
=
2\sum_{i=1}^{j-1}\chi_i+\chi_j.
$$
For notational convenience, define
$$
\Gamma_j(\bm\chi)
:=
2\sum_{i=1}^{j-1}\chi_i+\chi_j.
$$
Moreover, let
$$
w_j
:=
\int_{x_j}^{x_{j+1}}g(t)\,dt,
\quad
j=1,\ldots,N.
$$

For a $1$-Lipschitz function with prescribed endpoint values on $[x_j,x_{j+1}]$, the integral $w_j$ is bounded above and below by
$$
\frac{h_j}{2}
\left(
g(x_j)+g(x_{j+1})
\right)
\pm
\frac{h_j^2-\chi_j^2}{4}.
$$
Thus, the Kantorovich distance can be written as the following quadratically constrained problem:
\begin{subequations}
\label{eq:Kant-u-v-LP-new-x}
\begin{align}
\dd_K
\left(
u_{\X},
\tilde{u}_{\X}^{k-1}
\right)
=
\max_{\bm w,\bm\chi}
\quad
&
\sum_{j=1}^{N}
\left(
\beta_j-\tilde{\beta}_j^{k-1}
\right)w_j
\label{eq:kanto-qcqp-obj}
\\
\text{s.t.}\quad
&
w_j
\leq
\frac14h_j^2
-\frac14\chi_j^2
+\frac12h_j\Gamma_j(\bm\chi),
\;
j=1,\ldots,N,
\label{eq:kanto-qcqp-upper}
\\
&
-w_j
\leq
\frac14h_j^2
-\frac14\chi_j^2
-\frac12h_j\Gamma_j(\bm\chi),
\;
j=1,\ldots,N.
\label{eq:kanto-qcqp-lower}
\end{align}
\end{subequations}
The two inequalities in
\eqref{eq:kanto-qcqp-upper}--\eqref{eq:kanto-qcqp-lower}
also enforce $|\chi_j|\leq h_j$, which is precisely the endpoint condition implied by the $1$-Lipschitz property.

Define the three-dimensional second-order cone by
$$
\operatorname{SOC}_3
:=
\left\{
(y_1,y_2,y_3)\in\R^3
\ \middle|\
\sqrt{y_1^2+y_2^2}
\leq y_3
\right\}.
$$
Each quadratic constraint in
\eqref{eq:Kant-u-v-LP-new-x}
admits an equivalent second-order cone representation. Therefore,
\begin{subequations}
\label{eq:Kant-u-v-LP-new-socp}
\begin{align}
\dd_K
\left(
u_{\X},
\tilde{u}_{\X}^{k-1}
\right)
=
\max_{\bm w,\bm\chi}
\quad
&
\sum_{j=1}^{N}
\left(
\beta_j-\tilde{\beta}_j^{k-1}
\right)w_j
\label{eq:kanto-primal-socp-obj}
\\
\text{s.t.}\quad
&
\begin{bmatrix}
\chi_j
\\[1mm]
-2w_j
+h_j\Gamma_j(\bm\chi)
+\dfrac12h_j^2-\dfrac12
\\[2mm]
-2w_j
+h_j\Gamma_j(\bm\chi)
+\dfrac12h_j^2+\dfrac12
\end{bmatrix}
\in\operatorname{SOC}_3,\nonumber\\
&\qquad\qquad\qquad\text{for}\; j=1,\ldots,N,
\label{eq:kanto-primal-socp-upper}
\\
&
\begin{bmatrix}
\chi_j
\\[1mm]
2w_j
-h_j\Gamma_j(\bm\chi)
+\dfrac12h_j^2-\dfrac12
\\[2mm]
2w_j
-h_j\Gamma_j(\bm\chi)
+\dfrac12h_j^2+\dfrac12
\end{bmatrix}
\in\operatorname{SOC}_3,\nonumber\\
&\qquad\qquad\qquad\text{for}\; j=1,\ldots,N.
\label{eq:kanto-primal-socp-lower}
\end{align}
\end{subequations}

Problem \eqref{eq:Kant-u-v-LP-new-socp} satisfies Slater's condition. For example, $\bm w=\bm 0$ and $\bm\chi=\bm 0$ place both cone constraints in their interiors because $h_j>0$. Therefore, strong conic duality holds.

Let
$
\left(
\tau_j^+,\eta_j^+,\theta_j^+
\right)
\in\operatorname{SOC}_3$ and $\left(
\tau_j^-,\eta_j^-,\theta_j^-
\right)
\in\operatorname{SOC}_3
$
be the dual variables associated with
\eqref{eq:kanto-primal-socp-upper}
and
\eqref{eq:kanto-primal-socp-lower},
respectively. Since $\operatorname{SOC}_3$ is self-dual, the conic dual is
\begin{subequations}
\begin{align}
\dd_K&
\left(
u_{\X},
\tilde{u}_{\X}^{k-1}
\right)
=\\
\min\quad
&
\frac12
\sum_{j=1}^{N}
h_j^2
\left(
\eta_j^+
+\theta_j^+
+\eta_j^-
+\theta_j^-
\right)
-
\frac12
\sum_{j=1}^{N}
\left(
\eta_j^+
-\theta_j^+
+\eta_j^-
-\theta_j^-
\right)
\\
\text{s.t.}\quad
&
\frac12
\left(
\beta_j-\tilde{\beta}_j^{k-1}
\right)
-\eta_j^+
-\theta_j^+
+\eta_j^-
+\theta_j^-
=0,\;
j=1,\ldots,N,
\\
&
\tau_j^+
+\tau_j^-
+h_j
\left(
\eta_j^+
+\theta_j^+
-\eta_j^-
-\theta_j^-
\right)\\
&\quad+
2\sum_{i=j+1}^{N}
h_i
\left(
\eta_i^+
+\theta_i^+
-\eta_i^-
-\theta_i^-
\right)
=0, \;j=1,\ldots,N,
\\
&
\begin{bmatrix}
\tau_j^+\\
\eta_j^+\\
\theta_j^+
\end{bmatrix}
\in\operatorname{SOC}_3,
\quad
\begin{bmatrix}
\tau_j^-\\
\eta_j^-\\
\theta_j^-
\end{bmatrix}
\in\operatorname{SOC}_3,\;
j=1,\ldots,N.
\end{align}
\end{subequations}

Combining the conic dual representation of the Kantorovich distance with the preference and shape constraints yields the following SOCP reformulation of the piecewise-linear utility-updating problem:
\begin{subequations}
\label{eq:final-reformulation}
\begin{align}
\min_{\substack{
\boldsymbol{\alpha},\boldsymbol{\beta},\zeta,\\
\bm\tau^+,\bm\eta^+,\bm\theta^+,\\
\bm\tau^-,\bm\eta^-,\bm\theta^-
}}
\quad
&
\zeta
\\
\text{s.t.}\quad
&
\zeta
\geq
\frac12
\sum_{j=1}^{N}
h_j^2
\left(
\eta_j^+
+\theta_j^+
+\eta_j^-
+\theta_j^-
\right)
\nonumber\\
&
\qquad\qquad\qquad
-
\frac12
\sum_{j=1}^{N}
\left(
\eta_j^+
-\theta_j^+
+\eta_j^-
-\theta_j^-
\right),
\\
&
\frac12
\left(
\beta_j-\tilde{\beta}_j^{k-1}
\right)
-\eta_j^+
-\theta_j^+
+\eta_j^-
+\theta_j^-
=0,\;
j=1,\ldots,N,
\\
&
\tau_j^+
+\tau_j^-
+h_j
\left(
\eta_j^+
+\theta_j^+
-\eta_j^-
-\theta_j^-
\right)
\nonumber\\
&
+
2\sum_{i=j+1}^{N}
h_i
\left(
\eta_i^+
+\theta_i^+
-\eta_i^-
-\theta_i^-
\right)
=0,\;
j=1,\ldots,N, 
\\
&
\begin{bmatrix}
\tau_j^+\\
\eta_j^+\\
\theta_j^+
\end{bmatrix}
\in\operatorname{SOC}_3,
\;
\begin{bmatrix}
\tau_j^-\\
\eta_j^-\\
\theta_j^-
\end{bmatrix}
\in\operatorname{SOC}_3,\;
j=1,\ldots,N,
\\
&
\beta_j
=
\frac{\alpha_{j+1}-\alpha_j}{h_j},
\qquad
j=1,\ldots,N,
\label{con:alpha_beta}
\\
&
\sum_{i=1}^{n_A}
p_i^A u_{\X}(r_i^A)
\geq
\sum_{i=1}^{n_B}
p_i^B u_{\X}(r_i^B),\;
\forall (A,B)\in\Q^k,
\label{con:preference}
\\
&
\alpha_1
=
\tilde{\alpha}_1^{k-1},
\qquad
\alpha_{N+1}
=
\tilde{\alpha}_{N+1}^{k-1},
\label{con:end}
\\
&
0\leq\beta_j\leq L,
\qquad
j=1,\ldots,N.
\label{con:lip}
\end{align}
\end{subequations}


Constraint~\eqref{con:alpha_beta} links the breakpoint values and segment slopes, ensuring that they define the same continuous piecewise-linear utility function. Constraint~\eqref{con:preference} represents the pairwise preference constraints in \eqref{eq:updating_PLA_cons}. Since $u_{\X}$ is affine on each breakpoint interval, these constraints are linear in $\boldsymbol{\alpha}$ and $\boldsymbol{\beta}$, regardless of whether the product outcomes coincide with the breakpoints. Constraint \eqref{con:lip} imposes monotonicity and Lipschitz continuous. Constraint~\eqref{con:end} requires the updated utility $u_{\X}$ and the reference utility $\tilde{u}_{\X}^{k-1}$ to have identical values at the two endpoints of the current domain. This condition eliminates the boundary term in the integration-by-parts representation of the Kantorovich distance. Since the objective is linear, the Kantorovich-distance representation consists of second-order cone constraints, and all remaining constraints are linear, problem~\eqref{eq:final-reformulation} is an SOCP and can be solved using standard conic solvers.

It is worth noting that constraint~\eqref{con:end} guarantees the endpoint values are inherited from the linear extension of the previous nominal utility whenever the domain is expanded in Algorithm~\ref{alg:incvnm}.

\section{Direction error}
\label{apx:direction_error}

Consider two random-outcome products $A$ and $B$, and suppose that the user strictly prefers $B$ to $A$. Under von Neumann--Morgenstern expected utility theory, this preference is represented by
\begin{equation}
\label{eq:true_preference_margin}
\Delta_{u^*}(A,B)
:=
V_{u^*}(B)-V_{u^*}(A)
=
\mathbb E[u^*(B)]
-
\mathbb E[u^*(A)]
>0,
\end{equation}
where $u^*\in\U_L$ is the user's true utility function.

Let $\Pi_{\X}u^*$ denote the piecewise-linear approximation of $u^*$ over the breakpoint set
$
\X=\{x_1<\cdots<x_{N+1}\},
$
as defined in \eqref{eq:def_PLA}. The corresponding utility margin under the PLA is
\begin{equation}
\label{eq:pla_preference_margin}
\Delta_{\Pi_{\X}u^*}(A,B)
:=
V_{\Pi_{\X}u^*}(B)
-
V_{\Pi_{\X}u^*}(A).
\end{equation}

However, the PLA does not necessarily preserve the sign of the true utility margin. Indeed,
\begin{align*}
&
\Delta_{\Pi_{\X}u^*}(A,B)
-
\Delta_{u^*}(A,B)
\\
&=
\mathbb E
\left[
\bigl(\Pi_{\X}u^*-u^*\bigr)(B)
\right]
-
\mathbb E
\left[
\bigl(\Pi_{\X}u^*-u^*\bigr)(A)
\right].
\end{align*}
Hence, approximation errors at the outcomes of $A$ and $B$ may reverse the preference direction. We say that a \emph{direction error} occurs when
$$
\Delta_{u^*}(A,B)>0
\qquad
\text{but}
\qquad
\Delta_{\Pi_{\X}u^*}(A,B)\leq 0.
$$
This phenomenon is discussed in greater detail in
\citep{zhang2025modified,wei2026coordinate}.

No direction error occurs if all possible outcomes of $A$ and $B$ belong to the breakpoint set $\X$. Since the PLA coincides with the original utility function at every breakpoint,
$$
(\Pi_{\X}u^*)(x_i)
=
u^*(x_i),
\qquad
x_i\in\X,
$$
we have
\begin{equation*}
\mathbb E[(\Pi_{\X}u^*)(A)]
=
\mathbb E[u^*(A)],
\quad
\mathbb E[(\Pi_{\X}u^*)(B)]
=
\mathbb E[u^*(B)].
\end{equation*}
Therefore,
$$
\Delta_{\Pi_{\X}u^*}(A,B)
=
\Delta_{u^*}(A,B)
>0.
$$
Thus, the true and approximated preference directions coincide when the outcome supports of both products are contained in $\X$.

\section{Proofs}\label{apx:proofs}
\begin{proof}[\textbf{Proof of Proposition~\ref{prop:precision-error}}]
By the definition of nearest-grid rounding in
\eqref{eq:precision-rounding},
$$
\left|
r_i^A-\bar r_i^A
\right|
\leq
\frac{\varepsilon_{\mathrm{prec}}}{2},
\qquad
i=1,\ldots,n_A.
$$
Since $u\in\U_L$ is $L$-Lipschitz continuous,
$$
\left|
u(r_i^A)-u(\bar r_i^A)
\right|
\leq
L\left|
r_i^A-\bar r_i^A
\right|
\leq
\frac{L\varepsilon_{\mathrm{prec}}}{2}.
$$
Therefore,
\begin{align*}
\left|
V_u(A)-V_u(\bar A)
\right|
&=
\left|
\sum_{i=1}^{n_A}
p_i^A
\left[
u(r_i^A)-u(\bar r_i^A)
\right]
\right|
\\
&\leq
\sum_{i=1}^{n_A}
p_i^A
\left|
u(r_i^A)-u(\bar r_i^A)
\right|
\\
&\leq
\frac{L\varepsilon_{\mathrm{prec}}}{2}
\sum_{i=1}^{n_A}p_i^A
=
\frac{L\varepsilon_{\mathrm{prec}}}{2}.
    \qedhere
\end{align*}
\end{proof}

\begin{proof}[\textbf{Proof of Proposition~\ref{prop:scalable-update}}]
After the precision adjustment in \eqref{eq:precision-rounding}, every breakpoint belongs to the finite grid
$$
\mathcal G_{\varepsilon}
:=\X^0\cup\big(
[\underline{x},\overline{x}]
\cap
\varepsilon_{\mathrm{prec}}\mathbb Z
\big).
$$
where $\mathbb Z$ denotes the set of integers, so that
$\varepsilon_{\mathrm{prec}}\mathbb Z
:=\{j\varepsilon_{\mathrm{prec}}:j\in\mathbb Z\}$
is the uniform grid with spacing $\varepsilon_{\mathrm{prec}}$. The number of grid points satisfies
$$
|\mathcal G_{\varepsilon}|
\leq
|\X^0|+\left(\left\lfloor
\frac{\underline{x}-\overline{x}}
{\varepsilon_{\mathrm{prec}}}
\right\rfloor
+1\right)
:=
M_{\varepsilon_{\mathrm{prec}}}.
$$
Since the update in \eqref{eq:dynamic_breakpoint_precision} only adds elements of $\mathcal G_{\varepsilon}$, we have
$$
\X^k\subseteq\mathcal G_{\varepsilon}
$$
and hence $|\X^k|\leq M_{\varepsilon_{\mathrm{prec}}}$ for every $k$, which proves \eqref{eq:bp_set_bound}.

As shown in Appendix~\ref{apx:reform_updating}, the SOCP reformulation contains numbers of breakpoint values, segment slopes, conic auxiliary variables, and second-order cone constraints that are all linear in $|\X^k|$. A click sequence generated from a list of $n$ products induces
$n-1$ comparisons in $\Q^k$, where $n$ is a constant. Therefore, the hard-constraint utility
update has size
$
\mathcal O(|\X^k|).
$
And for the confidence-weighted update, the extended comparison set satisfies
\[
|\widetilde{\Q}^k|
=
\binom{T_k}{2}+n-T_k
\leq
\binom{n}{2}
\]
Since each comparison introduces one additional slack variable and one
relaxed preference constraint, the confidence-weighted update also has size
$
\mathcal O(|\X^k|).
$
Substituting
$
|\X^k|\leq M_{\varepsilon_{\mathrm{prec}}}
$
yields
$
\mathcal O(M_{\varepsilon_{\mathrm{prec}}})
$
for both the hard-constraint update and the confidence-weighted update. Both formulations remain uniformly bounded over the interaction history
and independent of the catalog size $|\cD|$.
\end{proof}

\section{Implementation details of the baseline methods}
\label{app:baseline_details}

This appendix describes the implementation of the baseline methods $
\mathcal M
=
\{\mathrm{ER},\mathrm{MV},\mathrm{DUCB},
\mathrm{DistBPR}\}
$. All methods use the settings stated in Section~\ref{sec:numerical}. At each interaction, each method assigns a scalar score $s_m^k(A)$ to every stock $A\in\cD$, where $m\in\mathcal M$.


\subsection{Expected return and mean--variance baseline}
\label{app:mean_variance}

 For stock $A\in\cD$, let
$
(r_1^{A,k},\ldots,r_{24}^{A,k})
$
denote the returns contained in the historical window available at interaction $k$. Define
$$
\widehat{\mathbb{E}}^{\,k}[A]
:=
\frac{1}{24}
\sum_{j=1}^{24}r_j^{A,k}
\quad
\text{and}\quad
\widehat{\operatorname{Var}}^{\,k}(A)
:=
\frac{1}{23}
\sum_{j=1}^{24}
\left(
r_j^{A,k}
-
\widehat{\mathbb{E}}^{\,k}[A]
\right)^2.
$$
The Expected return baseline evaluates each stock according to its empirical mean return only, i.e., $s^k_{\operatorname{ER}}(A):=\widehat{\mathbb{E}}^{\,k}[A]$.
The mean--variance baseline considers both empirical mean return and return variance
\citep{markowitz1952portfolio}:
\begin{equation}
\label{eq:MV_score}
s_{\mathrm{MV}}^k(A)
:=
\widehat{\mathbb{E}}^{\,k}[A]
-
\widehat{\operatorname{Var}}^{\,k}(A).
\end{equation}
At each interaction, the eight stocks with the highest scores form the next product list. 

\subsection{Discounted upper confidence bound method}
\label{app:discounted_ucb}

The discounted upper confidence bound method treats each stock as an arm and constructs product-level rewards from the observed click sequences. It adapts the UCB principle to a nonstationary multiple-play setting by discounting historical exposures and rewards
\citep{auer2002finite,garivier2011upper}.

\tightparagraph{Reward constructed from click sequences}
At interaction $k$, suppose that the displayed product list and the observed click sequence are
$
\Sl^k
=
(S_1^k,\ldots,S_8^k)
$
and
$
\C^k
=
(C_1^k,\ldots,C_{T^k}^k),
\;
T^k\leq 8,
$
respectively. It converts the ordinal click feedback into a normalized Borda reward. For each displayed stock $S\in\Sl^k$, define
\begin{equation}
\label{eq:ucb_borda_reward}
y_S^k
:=
\begin{cases}
\displaystyle
\frac{8-i}{7},
&
S=C_i^k,\quad i=1,\ldots,T^k,
\\[2mm]
0,
&
S\in
\Sl^k
\setminus
\{C_1^k,\ldots,C_{T^k}^k\},
\end{cases}
\end{equation}
as Borda reward. Stocks clicked earlier receive larger rewards, while unclicked stocks receive zero reward.

\tightparagraph{Discounted UCB score}
For each stock $A\in\cD$, let $N_A^k$ and $G_A^k$ denote its discounted effective exposure count and discounted cumulative reward after interaction $k$, respectively. They are initialized by
$
N_A^0=G_A^0=0,
$
and updated according to
\begin{subequations}
\label{eq:discounted_ucb_update}
\begin{align}
N_A^k
&=
\rho N_A^{k-1}
+
\mathbb{I}\{A\in\Sl^k\},
\label{eq:discounted_exposure_update}
\\
G_A^k
&=
\rho G_A^{k-1}
+
\mathbb{I}\{A\in\Sl^k\}y_A^k,
\label{eq:discounted_reward_update}
\end{align}
\end{subequations}
where $\rho\in(0,1]$ is the discount factor. When $N_A^k>0$, the discounted empirical reward is
\begin{equation}
\label{eq:discounted_empirical_reward}
\widehat q_A^k
:=
\frac{G_A^k}{N_A^k}.
\end{equation}

The DUCB score is defined as
\begin{subequations}
\label{eq:ucb_score}
\begin{empheq}[
left={s_{\mathrm{DUCB}}^k(A)=\empheqlbrace}
]{align}
&
\widehat q_A^k
+
\lambda
\sqrt{
\frac{
\log\left(
1+\sum_{B\in\cD}N_B^k
\right)
}{
N_A^k
}
},
&&
\text{if }N_A^k>0,
\label{eq:discounted_ucb_score}
\\
&
+\infty,
&&
\text{if }N_A^k=0,
\label{eq:ucb_unobserved_score}
\end{empheq}
\end{subequations}
where $\lambda\geq 0$ controls the strength of exploration. The first term favors stocks with high discounted historical rewards, while the second favors stocks with fewer effective exposures. Assigning an infinite score to an unobserved stock ensures that previously unseen stocks receive priority for exploration.

After interaction $k$, all stocks are ranked in descending order of $s_{\mathrm{DUCB}}^k(A)$, and the eight highest-ranked stocks form the product list for interaction $k+1$. All score ties, including those among unobserved stocks with infinite DUCB scores, are broken uniformly at random.

\begin{figure}[htbp]
    \centering
    \includegraphics[width=0.9\linewidth]{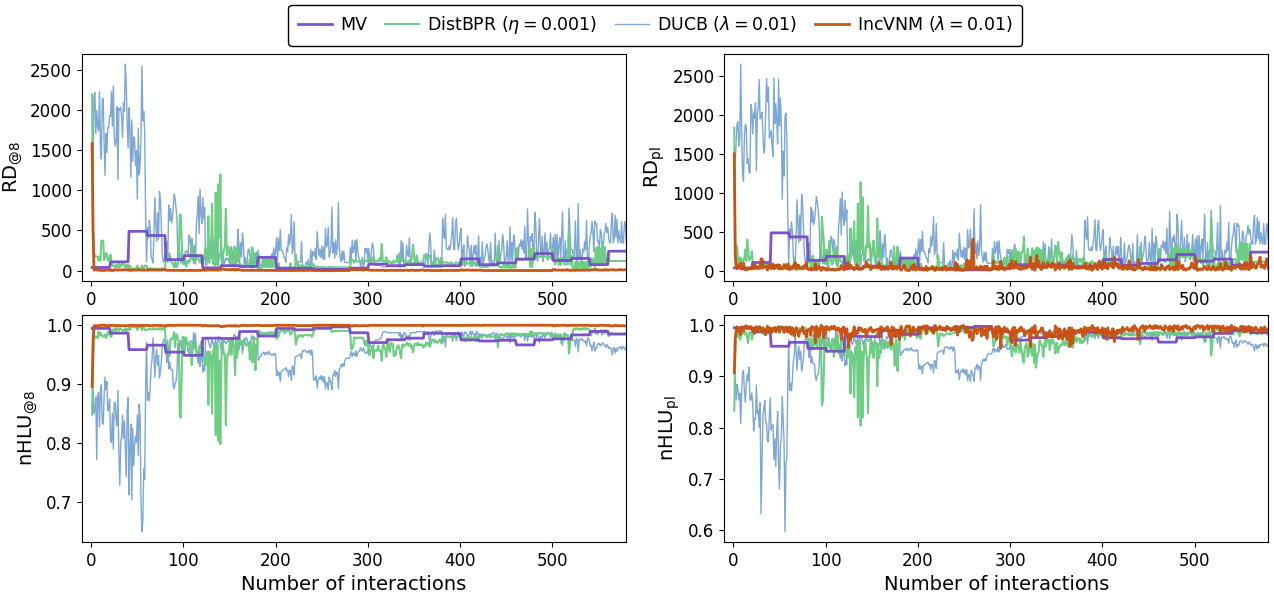}
    \caption{Performance evolution of utility learning and recommendation of methods.}
    \Description{Four plots compare the top-eight metrics and product list metrics over the interaction process.}
    \label{fig:baseline_comparison}
\end{figure}

\subsection{Distribution-aware Bayesian personalized ranking method}
\label{app:bpr_dist}

Traditional Bayesian personalized ranking (BPR) learns latent
user--item scores from historical interactions, making it less suitable
for products whose outcome distributions
change over time. We therefore replace the item representation in BPR
with features extracted from the current outcome distribution and refer
to the resulting baseline method as distribution-aware BPR (DistBPR).

\begin{table}[b]
\centering\caption{Cumulative-average ranking performance under error-free responses.}\label{tab:baseline_comparison}
\fontsize{8pt}{10pt}\selectfont\setlength{\tabcolsep}{2pt}
\renewcommand{\arraystretch}{0.82}
\begin{tabularx}{\textwidth}{@{}>{\raggedright\arraybackslash}p{2cm}>{\centering\arraybackslash}p{1.35cm}>{\centering\arraybackslash}p{0.70cm}*{11}{>{\centering\arraybackslash}X}@{}}
\toprule
\multirow{2}{*}{}&\multirow{2}{*}{}&\multirow{2}{*}{}&\multirow{2}{*}{ER}&\multirow{2}{*}{MV}&\multicolumn{3}{c}{DistBPR}&\multicolumn{3}{c}{DUCB}&\multicolumn{3}{c}{IncVNM}\\
\cmidrule(lr){6-8}\cmidrule(lr){9-11}\cmidrule(lr){12-14}&&&&&0.001&0.005&0.01&0.005&0.01&0.02&0.01&0.02&0.03\\
\midrule
\multirow{6}{1.8cm}{\centering Top-8 ranking}&\multirow{3}{*}{$\mathrm{RD}_{@8}\downarrow$}&200&567.90&170.30&155.22&126.84$^*$&137.38&735.01&729.76$^*$&758.20&15.63&13.37&\textbf{13.31$^*$}\\
&&400&409.10&107.35&122.69&121.20&113.28$^*$&492.42&467.52$^*$&505.38&8.90&\textbf{7.71$^*$}&8.00\\
&&580&378.86&117.48&119.92&112.88&109.52$^*$&476.46&419.64$^*$&457.13&7.26&\textbf{6.50$^*$}&6.84\\
\addlinespace[1pt]\cmidrule(lr){2-14}\addlinespace[1pt]
&\multirow{3}{*}{$\mathrm{nHLU}_{@8}\uparrow$}&200&0.9192&0.9736&0.9693&0.9724$^*$&0.9704&0.9166&0.9168$^*$&0.9148&0.9988&0.9989&\textbf{0.9990$^*$}\\
&&400&0.9433&0.9799&0.9727&0.9728&0.9738$^*$&0.9306&0.9371$^*$&0.9361&0.9993&\textbf{0.9994$^*$}&0.9994\\
&&580&0.9491&0.9792&0.9763&0.9764&0.9769$^*$&0.9410&0.9494$^*$&0.9468&0.9995&\textbf{0.9995$^*$}&0.9995\\
\midrule
\multirow{6}{1.55cm}{\centering Recommendation}&\multirow{3}{*}{$\mathrm{RD}_{\mathrm{pl}}\downarrow$}&200&567.90&170.30&158.74$^*$&161.34&190.91&750.52&745.27$^*$&773.71&\textbf{50.61$^*$}&95.79&102.97\\
&&400&409.10&107.35&132.94$^*$&138.24&145.86&500.18&475.27$^*$&513.13&\textbf{49.61$^*$}&88.58&105.21\\
&&580&378.86&117.48&131.16&129.02$^*$&136.21&481.81&424.99$^*$&462.48&\textbf{49.31$^*$}&91.92&108.99\\
\addlinespace[1pt]\cmidrule(lr){2-14}\addlinespace[1pt]
&\multirow{3}{*}{$\mathrm{nHLU}_{\mathrm{pl}}\uparrow$}&200&0.9192&0.9736&0.9667&0.9672$^*$&0.9625&0.9156&0.9158$^*$&0.9137&\textbf{0.9889$^*$}&0.9885&0.9748\\
&&400&0.9433&0.9799&0.9703$^*$&0.9695&0.9687&0.9301&0.9366$^*$&0.9356&\textbf{0.9884$^*$}&0.9884&0.9745\\
&&580&0.9491&0.9792&0.9744$^*$&0.9733&0.9729&0.9407&0.9490$^*$&0.9465&\textbf{0.9898$^*$}&0.9897&0.9770\\
\bottomrule
\end{tabularx}
\vspace{2pt}
\begin{minipage}{\textwidth}
\footnotesize
\textit{Notes.}
The symbols $\downarrow$ and $\uparrow$ indicate that smaller and larger
values, respectively, are preferred.
An asterisk ($^*$) indicates the best parameter setting within each
parameterized method family, while boldface indicates the best result
among all methods.
Best results are identified using the unrounded values.
\end{minipage}
\end{table}
\tightparagraph{Distributional product representation}
Let
$
\mathcal B
=
\{B_1,\ldots,B_d\}
$
be a fixed partition of the return range into $d$ disjoint intervals.
At interaction $k$, stock $A\in\cD$ has $24$
historical return observations
$
(r_1^{A,k},\ldots,r_{24}^{A,k})
$
in the current rolling window. Its distributional feature vector is
defined by the empirical histogram
\begin{equation*}
\label{eq:bpr_histogram}
\bm\phi^k(A)
:=
\left(
\phi_1^k(A),\ldots,\phi_d^k(A)
\right)^\top,
\quad
\phi_i^k(A)
:=
\frac{1}{24}
\sum_{j=1}^{24}
\mathbb I
\left\{
r_j^{A,k}\in B_i
\right\}.
\end{equation*}
Thus, $\bm\phi^k(A)$ provides a fixed-dimensional representation of the
entire empirical outcome distribution of stock $A$.

For a fixed parameter vector
$\bm w^k\in\mathbb R^d$, the DistBPR score of stock $A$ is
\begin{equation}
\label{eq:bpr_dist_score}
s_{\mathrm{DistBPR}}^k(A)
:=
(\bm w^k)^\top \bm\phi^k(A).
\end{equation}

\tightparagraph{Incremental BPR update}
At interaction $k$, the displayed product list $\Sl^k$ and the observed
click sequence $\C^k$ generate the pairwise comparison set
$
\Q^k
=
\{(A_q,B_q)\}.
$
DistBPR models the corresponding preference probability as
\begin{equation}
\label{eq:bpr_probability}
\mathbb P
\left(
A_q\succeq B_q
\right)
=
\sigma
\left(
\bm w^\top
\left[
\bm\phi^k(A_q)-\bm\phi^k(B_q)
\right]
\right),
\end{equation}
where
$\sigma(z):=1/(1+\exp(-z))$ is the sigmoid function. With response errors, 
$\Q^k$ is replaced with $\tilde{\Q}^k$.

The parameter vector is incrementally updated by solving
\begin{equation*}
\label{eq:bpr_dist_update}
\bm w^k
\in
\arg\min_{\bm w\in\mathbb R^d}
\left\{
-
\sum_{(A,B)\in\Q^k}
\log
\sigma
\left(
\bm w^\top
\left[
\bm\phi^k(A)-\bm\phi^k(B)
\right]
\right)
+
\eta
\left\|
\bm w-\bm w^{k-1}
\right\|_2^2
\right\},
\end{equation*}
where $\eta>0$ controls the extent to which the updated model retains
the parameter estimate from the previous interaction and $\bm w^0$ is initialized with $0$.
The first term is the standard BPR
pairwise-ranking loss, while the second term supports incremental
learning by preventing abrupt changes in the scoring model. 

\tightparagraph{Product-list generation}
To ensure a fair comparison with IncVNM, DistBPR uses the same
entropy-regularized sampling policy for product-list generation. After
interaction $k$, the sampling probability assigned to stock
$A\in\cD$ is
\begin{equation}
\label{eq:bpr_dist_sampling}
\mu_A^k
=
\frac{
\exp
\left(
s_{\mathrm{DistBPR}}^k(A)/\lambda_{\mathrm{Dist}}
\right)
}{
\displaystyle
\sum_{B\in\cD}
\exp
\left(
s_{\mathrm{DistBPR}}^k(B)/\lambda_{\mathrm{Dist}}
\right)
},
\end{equation}
where $\lambda_{\mathrm{Dist}}>0$ controls the exploration strength, which is set to be 0.02 in our experiment. The product list
for interaction $k+1$ is obtained by sampling eight distinct stocks according to $\bm\mu^k$.
At the first interaction, $\bm w^0=\bm 0$ gives identical scores to all
stocks, and the initial product list is therefore selected
uniformly at random. 

\section{Results under error-free responses}
\label{sec:results-errorfree}

We test an error-free user whose click sequences are fully consistent with the true utility function $u^*$. 
Accordingly, IncVNM uses the hard-constraint utility update in \eqref{eq:updating_PLA_flexible}. 
Table~\ref{tab:baseline_comparison} reports the cumulative averages of $\operatorname{RD}_{@8}$, $\operatorname{nHLU}_{@8}$, $\mathrm{RD}_{\mathrm{pl}}$ and $\operatorname{nHLU}_{\mathrm{pl}}$ after 200, 400, and 580 interactions. Figure~\ref{fig:baseline_comparison} presents the interaction-level trajectories for MV, DistBPR with $\eta=0.001$, IncVNM and DUCB with $\lambda=0.01$. Each entry is the
mean of the corresponding cumulative-average metric over 15 runs,
comprising five repetitions for each of three random seeds. The results lead to three main observations.

\noindent\textbf{(a) IncVNM achieves the best overall performance in both utility learning and recommendation.}
    Its utility-updating metrics are consistently and substantially better than those of all baselines, with stable performance throughout the interaction process. The recommendation metrics exhibit occasional fluctuations because IncVNM continues to explore new products, but remains clearly superior overall.

    \noindent\textbf{(b) DistBPR, DUCB, and IncVNM all exhibit learning effects as the number of interactions increases.}
     DUCB learns product-level ranking, which limits its ability to identify preferences over products with random outcomes. DistBPR learns a distribution-level ranking score from click feedback, but it does not explicitly model the user's attitude toward risk. In contrast, MV does not learn from clicks, hence there is no decreasing trend as the number of interactions increases. But its variance penalty partially captures the simulated user's risk aversion and therefore yields relatively competitive performance.

    \noindent\textbf{(c) The exploration rate of IncVNM reflects a clear trade-off between utility learning and immediate recommendation quality.}
    A larger $\lambda$ generally provides more informative exploration and slightly improves utility updating, whereas a smaller $\lambda$ generates product lists that better match the user's current preferences. This contrast is particularly visible between $\lambda=0.01$ and $\lambda=0.02$. Nevertheless, IncVNM remains the best-performing method under all tested settings, indicating that its overall advantage is robust to moderate parameter variation.

\section{Experimental settings}
\label{app:experimental_settings}

\begin{table}[!htbp]
\centering
\caption{Hyperparameters, initialization, and implementation settings.}
\label{tab:experimental_settings}
\small
\setlength{\tabcolsep}{5pt}
\renewcommand{\arraystretch}{1.1}
\begin{tabularx}{\textwidth}{
@{}
>{\raggedright\arraybackslash}p{2.2cm}
>{\raggedright\arraybackslash}p{3.4cm}
>{\raggedright\arraybackslash}X
@{}}
\toprule
Component & Parameter & Setting \\
\midrule

\multirow{9}{*}{IncVNM}
& Product-list size & $n=8$ \\

& Historical window length & 24 months \\

& Numerical precision
& $\varepsilon_{\mathrm{prec}}=0.01$ \\

& Initial breakpoint set
& $\X^0=\{-0.2,0.2\}$ \\

& Initial nominal utility
& $\tilde u_{\X^0}^{0}(-0.2)=0$ and
  $\tilde u_{\X^0}^{0}(0.2)=1$ \\

& Lipschitz modulus
& $L=100$ \\

& Confidence-scale parameter
& $m=100$ \\

& Violation-penalty parameter
& $\gamma=3$ \\

& Conic solver
& \textsc{Clarabel} \\

\midrule

\multirow{1}{*}{DUCB}
& Discount factor
& $\rho=0.995$ \\



\midrule

\multirow{6}{*}{DistBPR}
& Histogram dimension
& $d=10$ \\

& Histogram calibration data
& All stock returns in the initial 24-month window \\

& Histogram range
& Let $r_{\min}^0$ and $r_{\max}^0$ be the minimum and maximum
  calibration returns and let
  $R^0=r_{\max}^0-r_{\min}^0$.
  The calibrated range is
  $[r_{\min}^0-0.05R^0,\,
    r_{\max}^0+0.05R^0]$ \\

& Bin construction
& Ten equal-width bins fixed throughout the experiment;
  subsequent observations outside the calibrated range are assigned
  to the corresponding boundary bins \\

& Initial parameter vector
& $\bm w^0=\bm 0\in\mathbb R^{10}$ \\


& Sampling parameter
& $\lambda_{\mathrm{Dist}}=0.02$ \\


\midrule

\multirow{2}{*}{Evaluation}
& Half-life parameter
& $h=4$ \\

& Stabilization constant
& $\epsilon=10^{-12}$ \\

\midrule

\multirow{2}{*}{Simulation}
& True utility
& $u^*(x)=1-0.8\exp(-6x)$ \\

& Click threshold
& Median oracle expected utility over the current catalog,
  corresponding to quantile $0.5$ \\

\bottomrule
\end{tabularx}
\end{table}

\end{document}